\documentclass[usenatbib,usegraphicx]{mn2e}

\usepackage{aas_macros}
\usepackage{verbatim}
\usepackage{amsmath}
\usepackage{amsfonts}
\usepackage{amssymb,color}
\usepackage{empheq}
\usepackage[charter]{mathdesign}
\usepackage{charter}
\usepackage{url}
\bibpunct[, ]{(}{)}{;}{a}{}{,}
\newcommand{\spin}{\ensuremath{\tau_*}}
\newcommand{\spinf}{\ensuremath{\nu_*}}
\newcommand{\spotf}{\ensuremath{\nu_s}}
\newcommand{\spott}{\ensuremath{\tau_s}}
\newcommand{\tnext}{\ensuremath{t^{next}}}
\newcommand{\tnextmin}{\ensuremath{t^{next}_{min}}}
\newcommand{\tnextmax}{\ensuremath{t^{next}_{max}}}
\newcommand{\tdur}{\ensuremath{\Delta T}}
\newcommand{\tdurmin}{\ensuremath{(\Delta T)_{min}}}
\newcommand{\tdurmax}{\ensuremath{(\Delta T)_{max}}}
\newcommand{\vel}{\ensuremath{\mathbf{v}}}
\newcommand{\R}{\ensuremath{\mathbf{R}}}
\newcommand{\Om}{\ensuremath{\mathbf{\Omega}}}
\newcommand{\magn}{\ensuremath{\mathbf{B}}}
\newcommand{\g}{\ensuremath{\mathbf{g}}}
\newcommand{\mfluxden}{\ensuremath{\mathcal{T}}}
\newcommand{\entropy}{\ensuremath{{S}}}

\newcommand{\Ps}{\ensuremath{\mathcal{P}}}
\newcommand{\Ws}{\ensuremath{\mathcal{W}}}
\newcommand{\U}{\ensuremath{\mathcal{U}}}
\newcommand{\mdot}{\ensuremath{\dot{M}}}
\newcommand{\adim}{\widetilde}
\newcommand{\amdot}{\ensuremath{\adim{\mdot}}}
\newcommand{\amu}{\ensuremath{\adim{\mu}}}
\newcommand{\aL}{\ensuremath{\adim{L}}}
\newcommand{\ar}{\ensuremath{\adim{r}}}
\newcommand{\ac}{\ensuremath{\ar_c}}
\newcommand{\aalpha}{\ensuremath{\adim{\alpha}}}
\newcommand{\nuq}{\ensuremath{\nu_{QPO}}}
\newcommand{\mis}{\ensuremath{\Theta}}
\newcommand{\msun}{\ensuremath{M_{\odot}}}
\newcommand{\fref}{Fig.~\ref}
\newcommand{\eref}{Eq.~\ref}
\newcommand{\sref}{\S\ref}

\newcommand{\degree}{\ensuremath{^{\circ}}}

\makeatletter
\def\url@leostyle{%
  \@ifundefined{selectfont}{\def\UrlFont{\sf}}{\def\UrlFont{\small\ttfamily}}}
\makeatother
\title{QPO emission from moving hot spots on the surface of neutron stars: a model}
\author [M.Bachetti et al.]
{ Matteo Bachetti$^1$\thanks{mbachett@ca.astro.it},
  Marina M. Romanova$^2$\thanks{romanova@astro.cornell.edu}, 
  Akshay Kulkarni$^2$,
\newauthor
  Luciano Burderi$^1$ 
  and Tiziana di Salvo$^3$ \\
  $^1$Universit\`a degli Studi di Cagliari, Dipartimento di Fisica, SP Monserrato - Sestu, km 0.7, 09042 Monserrato, Italy\\
  $^2$Dept. of Astronomy, Cornell University, 14853 Ithaca NY, USA \\
  $^3$Dipartimento di Scienze Fisiche ed Astronomiche, Universit\`a di Palermo, via Archirafi n.36, 90123 Palermo, Italy 
}
\date{}
\begin{document}

\maketitle

\begin{abstract}

We present recent results of 3D magnetohydrodynamic simulations of neutron stars with small misalignment angles, as regards the features in lightcurves produced by regular movements of the hot spots during accretion onto the star. In particular, we show that the variation of position of the hot spot created by the infalling matter, as observed in 3D simulations, can produce high frequency Quasi Periodic Oscillations with frequencies associated with the inner zone of the disk. 
Previously reported simulations showed that the usual assumption of a fixed hot spot near the polar region is valid only for misalignment angles $\mis$ relatively large. Otherwise, two phenomena challenge the assumption: one is the presence of Rayleigh-Taylor instabilities at the disk-magnetospheric boundary, which produce tongues of accreting matter that can reach the star almost anywhere between the equator and the polar region; the other one is the motion of the hot spot around the magnetic pole during stable accretion. In this paper we start by showing that both phenomena are capable of producing short-term oscillations in the lightcurves. We then use Monte Carlo techniques to produce model lightcurves based on the features of the movements observed, and we show that the main features of kHz QPOs can be reproduced. Finally, we show the behavior of the frequencies of the moving spots as the mass accretion rate changes, and propose a mechanism for the production of double QPO peaks.
\end{abstract}


\section{Introduction}\label{sec:intro}

Quasi Periodic Oscillations (QPO) have been observed in many X-ray sources for the last 24 years \citep{vanderKlis:1985p2318,Lamb:1985p2327}. They appear as broad peaks in the power spectra, with a behavior dependent on the source state, the mass accretion rate, and other physical properties of the sources. They are present in both neutron star (hereafter NS) and black hole (BH) sources, sometimes with peculiar similarities between the two classes of objects \citep[see][for a review]{2006csxs.book...39V}. 

Of particular interest are the kHz QPOs: they were first discovered in the Low Mass X-ray Binary (LMXB) and Z source Sco-X1 \citep{1996ApJ...469L...1V}, to be then observed in the spectra of nearly all Z and atoll sources, as well as several weak LMXBs \citep[see for example][]{Strohmayer:1996p2317,Wijnands:1997}. 
They have frequencies between 300Hz and 1.2kHz \citep[see][]{2006csxs.book...39V}. The peaks move with timescales of hours/days and their quality factor $Q=\nu/\Delta\nu$ (ratio between the frequency and the width of the peak, a measure of coherence) is not constant, but can be related with the QPO frequency \citep{Mendez:2001p4640, diSalvo:2003p5128,Barret:2006p7645} and with the mass accretion rate \citep[see for example][]{Mendez:2006p5133}.

The kHz QPOs usually appear as a pair of peaks (the {\em twin peaks}), labeled ``upper'' and ``lower'' (and their frequencies, accordingly, $\nu_u$ and $\nu_l$).  While on timescales of hours the two frequencies change, their difference $\Delta\nu=\nu_u-\nu_l$ tends to be almost costant. As with the discovery of burst oscillations \citep{Strohmayer:1996p2317} and Accreting Millisecond X-ray Pulsars (hereafter AMXP) \citep {Wijnands:1998p1343} the spin frequencies $\spinf$ of more and more accreting neutron stars were measured, an interesting behavior was observed: $\Delta\nu$ was usually found to be close to either $\spinf$ or $0.5\spinf$. 
Recent works \citep[see for example][]{Mendez:2007p5756} call this assumption into question. In particular, it can be shown that $\Delta \nu$ could in principle be independent from $\spinf$.

It has been shown that the kHz QPOs frequency is correlated, at least on a given system and for short timescales, with the X-ray count rate \citep{1999ApJ...511L..49M} and thus also, probably, with the mass accretion rate. 

Many models have been developed to explain the kHz QPO phenomenon. They involve a number of different mechanisms, like for example relativistic precession frequencies \citep{Stella:1998p4748}, relativistic resonances \citep{Kluzniak:2001p4749}, beats between the keplerian frequency at a particular radius in the disk and the frequency of the star \citep{Alpar:1985p2315,Lamb:1985p2327,Miller:1998p4107,Lamb:2004p6829}, oscillations in the comptonizing medium \citep{Lee:1998p4811}. Recently, \citet{Lovelace:2007p4213} and \citet{Lovelace:2009p4323}  found and analyzed a radially localized Rossby wave instability which exists in the inner region of a disk around a magnetized star where the angular rotation rate of the matter decreases approaching the star.  This instability may explain the twin kilo-Hertz QPOs observed in some LMXBs. The importance for QPOs of the disk angular rotation rate having a maximum was discussed earlier by \citet{Alpar:2005,Alpar:2008}

Most of the models of kHz QPOs, based on the fact that QPOs also appear in black hole systems and there seems to be a link between $\nu_l$ with frequencies observed in black hole systems \citep{Psaltis:1999p4878}, try to explain the behavior of kHz QPOs in both the neutron star and the black hole cases, excluding the possibility of a surface emission. \citet{2005AN....326..812G} claim that, according to spectral emission properties of QPOs, in Atoll and Z sources periodic and quasi-periodic emission is more likely to be produced on the surface of the star.

During the past years MHD simulations have been amongst the most fruitful methods to investigate accreting magnetized stars \citep[see for example][]{Goodson:1997p5065, Romanova:2002p4187, Romanova:2003, Romanova:2004p4095, Kulkarni:2005p4093, Bessolaz:2008p4909, Long:2008p4237, Romanova:2008}. 3D simulations have given insight into the way matter accretes, showing in more detail trajectories, shapes of magnetic field lines, shapes and movements of hot spots on the surface of the star \citep[see for example][]{Romanova:2003,Romanova:2004p4095}.

Simulations show that matter may accrete in several different regimes: 
\begin{itemize}
 \item the {\em stable accretion} regime, when matter is stopped by the magnetic field and conveyed to the magnetic poles via the so called {\em funnel flow} \citep{Romanova:2002p4187,Romanova:2003,Kulkarni:2005p4093}; 
 \item the {\em unstable} regime, when Rayleigh-Taylor instabilities form at the boundary between the disk and the magnetosphere \citep{Romanova:2006p2479,Romanova:2008,Kulkarni:2008}, and accretion takes place through tongues of matter from the inner edge of the disk to random places between the equator and the magnetic poles; 
 \item the {\em magnetic boundary layer} regime, during which the disk is very close to the surface of the star and the accretion takes place via two big antipodal equatorial tongues which rotate with the frequency of the inner disk \citep{Romanova:2009p6603}.
\end{itemize}

One of the interesting phenomena observed in 3D simulations is the movement of spots around the magnetic pole of the accreting star, both in stable \citep[][]{Romanova:2003,Romanova:2004p4095,Romanova:2006p3862} and in unstable regime \citep{Kulkarni:2009}. It was noticed that in some cases, such as in the magnetic boundary layer regime, the moving spots  produce QPO features \citep{Romanova:2009p6603}. Usually, models of NS surface emission (like in the case of Accreting Millisecond Pulsars) consider funneling of matter onto the polar caps as a static process, with a fixed polar cap and thus a fixed emission spot \\citep[e.g.][]{Beloborodov:2002p4173,Poutanen:2003p4169}. 3D simulations show that this may not be the case in particular for small misalignment angles.

Recent works \citep{Lamb:2009A,Lamb:2009B} investigate the wandering of hot spots at which matter accretes in sources with a small misalignment angle, finding that it might affect the coherence of the pulsed signal from neutron stars and even destroy it, being the origin of the transient emission in AMXPs. The movements hypothesized in those works, though, are different from the movements observed in our 3D simulations, where the motion of the hotspots is a rotational motion around the magnetic pole or in the equatorial region, on timescales of milliseconds. Moreover, the dimension of the hotspot in those two works (a radius of $\sim 25^{\circ}$) does not find confirmation in our simulations. It corresponds to the dimension of the whole ``hot ring'' where our hotspots (much smaller) move (see \fref{fig:sf-stab-join}).

In this paper we show that the observed movements carry with them the information of disk frequencies near the inner disk, and their effect is the production of quasi-periodic oscillations at the relative frequencies. We perform a series of 3D MHD simulations (mostly for the cases of very small $\mis$) and analyze the rotation of the spots and the related frequencies in detail. MHD simulations can only follow the motion of the disk on very short timescales compared to the disk evolution timescale and the observational time. For this reason, we use the features of the moving spots produced in simulations to generate long lightcurves by means of Monte Carlo techniques, showing that the motion observed produces QPOs. Then, we show that the phenomenon follows a predictable behavior as the mass accretion rate changes.

We start in \sref{sec:model} with a general description of the 3D MHD model used for this work. Then, we show the way spots move in our simulations in \sref{sec:3d}. We then show with a Monte Carlo-generated lightcurve our model of surface QPO production in \sref{sec:mc}. Finally, in \sref{sec:mdot} we show how the angular velocities of hot spots correlate with the mass accretion rate.

\section{Model used for 3D simulations}\label{sec:model}
We model the accretion disk around a neutron star as a single fluid plasma, interacting with the star's magnetic field. We then solve the complete system of MHD equations in the coordinate system rotating with the star \citep{Koldoba:2002p4178}
\begin{eqnarray}
 \frac{\partial\rho}{\partial t} + \nabla\cdot\left(\rho \vel\right) &=& 0\label{eq:cont-rho} \\
 \frac{\partial(\rho \vel)}{\partial t} + \nabla\cdot\mfluxden &=& \rho\g + 2(\rho\vel)\times\Om-\rho\Om\times(\Om\times\R)\label{eq:force} \\
 \frac{\partial\magn}{\partial t} &=&\nabla\times(\vel\times\magn) \label{eq:maxw} \\
 \frac{\partial(\rho\entropy)}{\partial t} +\nabla\cdot(\rho\entropy \vel)&=& 0 \label{eq:cont-entr},
\end{eqnarray}
where $\rho$ is the density, $\vel$ is the velocity of the fluid, $\g$ is the gravitational acceleration, $\magn$ is the magnetic field, $\mfluxden$ is the momentum flux-density tensor, and $\entropy$ is the density of entropy. The equations are solved by means of the Godunov scheme \citep{MR1998051}, using a cubed-sphere grid. The outward transfer of angular momentum in the disk is produced by an $\alpha$-type viscosity  \citep{Shakura:1973p4092}. A detailed description of these equations and their implementation can be found in \citet{Koldoba:2002p4178}.

The potential well around the neutron star is represented by the pseudo-Newtonian potential described in \citet{Paczynski}, in order to take into account relativistic corrections to the movement of matter around the star. This potential is described by the equation
\begin{equation}
 V(r) = - \frac{GM}{r-R_g}
\end{equation}
where $R_g = 2GM/c^2$ is Gravitational Radius (equal to the Schwarzschild Radius). This form of the potential gives the correct values for the marginally bound circular orbit and in particular the innermost stable circular orbit (ISCO) in the Schwarzschild non-rotating BH case. This can be useful to interpret the results of this work in terms of theoretical models taking the ISCO as a fundamental quantity for understanding QPOs.

\subsection{Conventions and units}\label{sec:units}
All quantities used in the 3D code are dimensionless. Hereafter, given a dimensional quantity $\U$, we call $\adim{\U}=\U/\U_0$ the corresponding dimensionless unit, where $\U_0$ is the reference value for $\U$. Every result can be applied to different physical systems, evaluating the $\U_0$s by choosing the right values of three fundamental dimensional parameters: 
\begin{itemize}
 \item $M$: the {\bf mass} of the star;
 \item $R$: its {\bf radius};
 \item $\mu$: its {\bf magnetic moment}, corresponding to a dipolar surface magnetic field $B=\mu/R^3$.
\end{itemize}
At the start of simulations, we fix the following dimensionless parameters:
\begin{itemize}
\item the {\bf corotation radius} $\ac$, defined as the radius where the Keplerian angular velocity equals the star's angular velocity. It sets the rotational velocity of the star (see \eref{eq:omega}); 
\item the {\bf viscosity} $\aalpha$, on the model of \citet{Shakura:1973p4092};
\item the {\bf magnetic moment} $\amu$, and possible non-dipolar components, which are not used in this work. Alternatively, $\amu$ can be considered as a parameter to set the size of the magnetosphere and the mass accretion rate (see \eref{eq:mdot});
\item the {\bf misalignment angle} $\mis$, the angle between the magnetic axis and the rotation axis.
\end{itemize}
The reference quantities are calculated as follows:
\begin{itemize}
 \item $M_0=M$;
 \item $r_0=R/0.35$, the reference {\bf distance} from the center of the star, chosen this way for convenience, because of the vicinity of $\ar=1$ to the usual range of magnetospheric radii;
 \item $\mu_0=\mu/\amu$ and $B_0=\mu_0/r_0^3$; then, the surface magnetic field is $B_*=(0.35)^{-3} B_0\amu$;
 \item $v_0=(GM/r_0)^{1/2}$, the reference {\bf velocity};
 \item $\omega_0=(GM/r_0^3)^{1/2}$, the reference {\bf angular velocity};
 \item $P_0=2\pi r_0/v_0$, the Keplerian period at $\ar=1$ as the reference {\bf time};
 \item $\rho_0=B_0^2/v_0^2$ as the reference {\bf density};
 \item $\mdot_0 = \rho_0 v_0 r_0^2$ as the reference {\bf mass accretion rate};
 \item $L_0 = \mdot_0 v_0^2$ is the reference {\bf luminosity}
\end{itemize}
from which we obtain 
\begin{eqnarray}
\spinf &=& \frac{(0.35)^{3/2}}{2\pi} \frac{(GM)^{1/2}}{(\ac R)^{3/2}}\label{eq:omega}\\
\spin &=& \frac{1}{\spinf}\label{eq:period}\\
\mdot &=& \mdot_0\amdot= \frac{(0.35)^{7/2}}{(GM)^{1/2}} \left(\frac{B_*}{\amu}\right)^2 R^{5/2} \amdot\label{eq:mdot}\\
L     &=& L_0\aL = (0.35)^{9/2}(GM)^{1/2}\left(\frac{B_*}{\amu}\right)^2 R^{3/2} \aL\label{eq:lum} 
\end{eqnarray}
where $\spinf$, $\spin$, $\mdot$ and $L$ are the frequency of the star, its period, the mass accretion rate and luminosity respectively. $\amdot$ and $\aL$ are returned by the code at every time step. $\aL$ is calculated under the assumption that all the energy of the accreted matter is converted into luminosity. Equation~\ref{eq:mdot} gives us a way to use the parameter $\amu$ to effectively measure the accretion rate. Instead of using the simplest interpretation, according to which $\amu$ is a dimensionless measure of the magnetic field, we can in fact change $B_0$ in order to make $B_*$ constant, changing effectively the mass accretion rate reference value $\mdot_0$.

\subsection{General information about the simulations in this paper}
Unless otherwise specified, throughout this paper all results are referred to a neutron star with $R=10$km, $M=1.4M_{\odot}$, $\mis=2\degree$. The magnetic field will be either $B=10^8$G or $B=5\cdot10^8$G. With these parameters, the gravitational radius of the Paczynski-Wiita potential is $R_g\approx4.16$km while the ISCO is at a radius $R_{ISCO}\approx12.5$km (in dimensionless units, $\ar=0.435$).

The grid used for most of the simulations has 72 cells in radial direction and 31 along every side of the six blocks of the cubed sphere. The simulations shown have been produced in the period January - June 2009, using the High Performance Computing facilities of NASA (Pleiades, Columbia) and CINECA (BCX).

\section{3d simulations and spot movements}\label{sec:3d}
In this section we show how spot motions can produce visible features in the lightcurves. 

\subsection{Stable, unstable, strongly unstable regimes}\label{sec:stab-unstab}

\begin{figure*}
 \centering
$\begin{array}{cc}
 \includegraphics[width=1.5in]{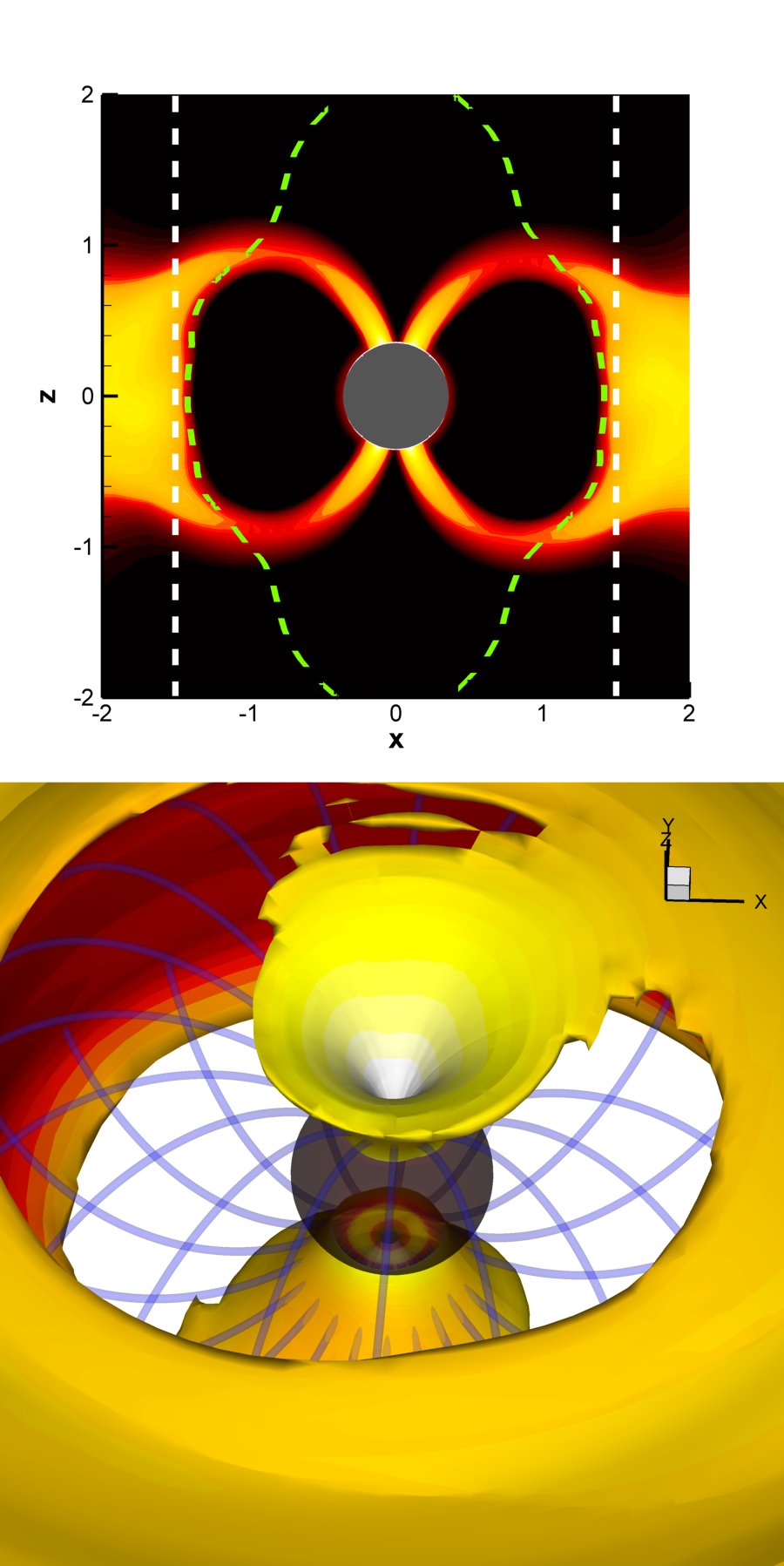}&
 \includegraphics[width=4.5in]{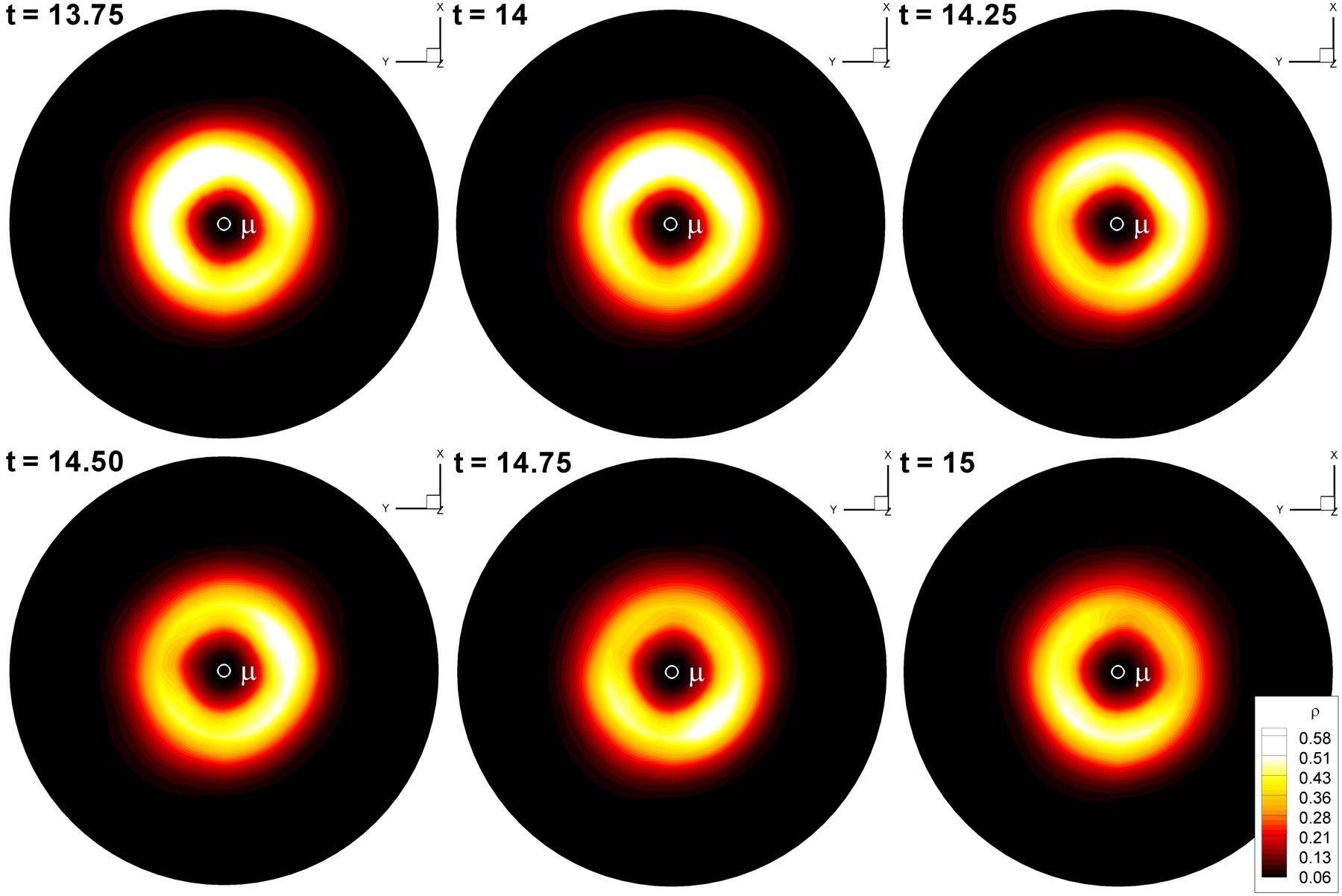}
\end{array}$
 \caption{Evolution of the position of the hot spot during accretion, for the case $\ac=1.5$ ($\spin=4.1$ms),  $\aalpha=0.04$, $\amu=2$, a good example of stable accretion. The time difference between the files is $0.5 P_0$. In this case all accretion takes place around the magnetic pole.}
 \label{fig:sf-stab-join}
\end{figure*}

\begin{figure*}
 \centering
$\begin{array}{cc}
 \includegraphics[width=1.5in]{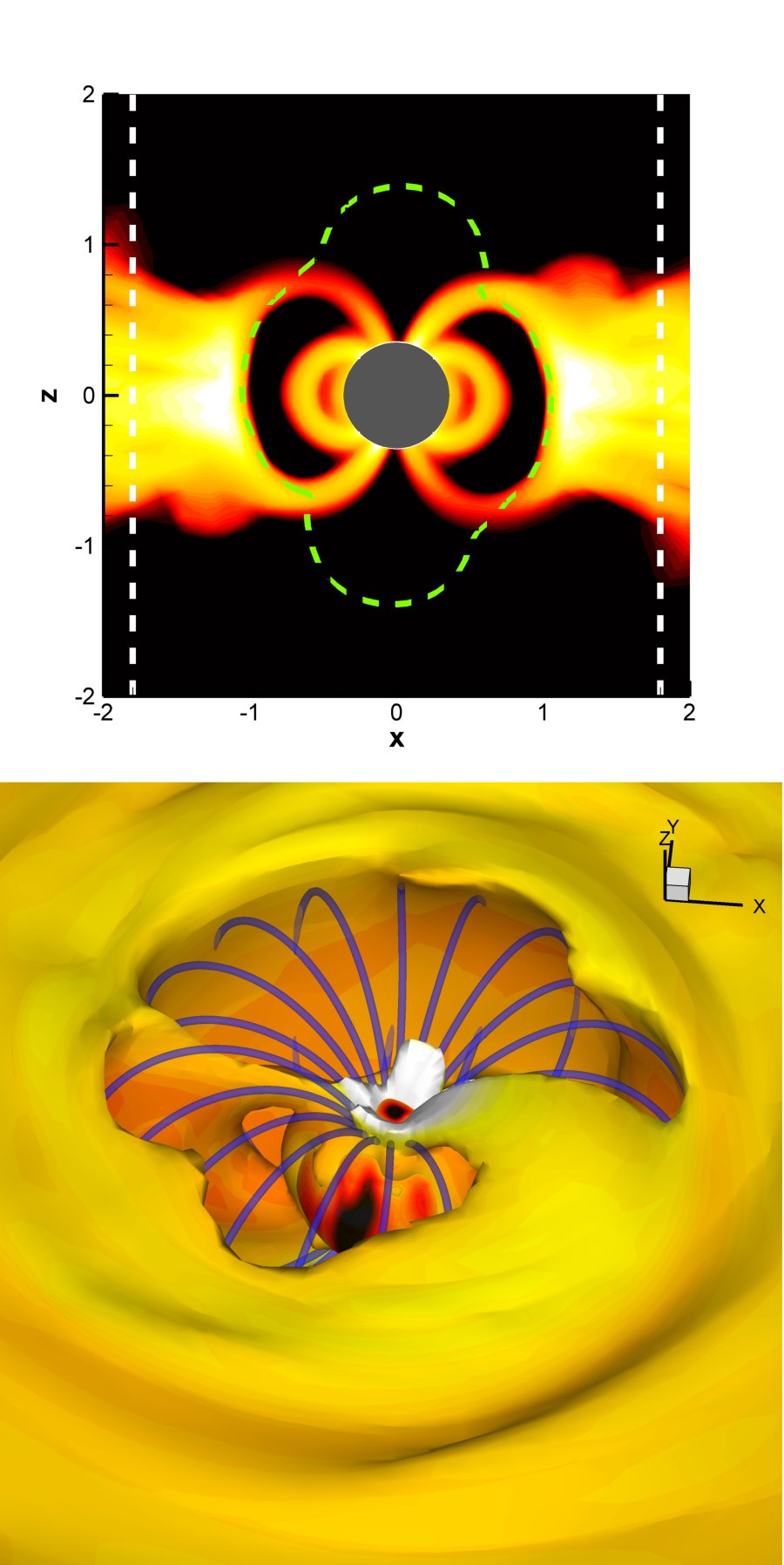}&
 \includegraphics[width=4.5in]{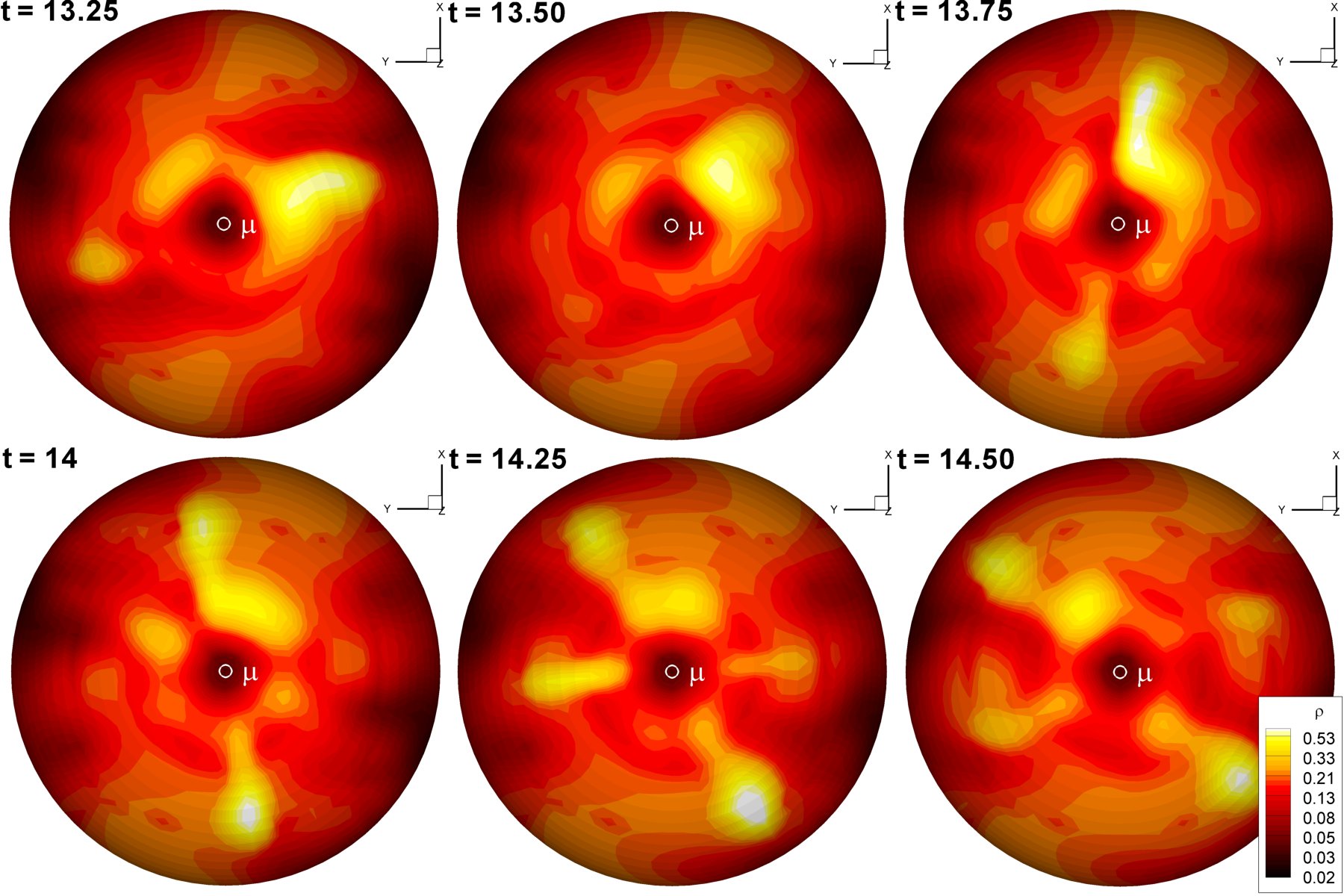}
\end{array}$
 \caption{Evolution of the position of the hot spot during accretion, for the case $\ac=1.8$ ($\spin=5.4$ms),  $\aalpha=0.03$, $\amu=0.9$, example of unstable accretion. The time difference between the files is $0.25 P_0$. In this case matter pressure is such that the magnetic field is not able to convey all accreting matter to the magnetic poles. Instead, instabilities form at the truncation radius, creating tongues of accreting matter which fall onto the surface in random places. Accretion takes place both in the polar region and in the rest of the surface}
 \label{fig:sf-unstab-join}
\end{figure*}
\begin{figure*}
 \centering
$\begin{array}{cc}
 \includegraphics[width=1.5in]{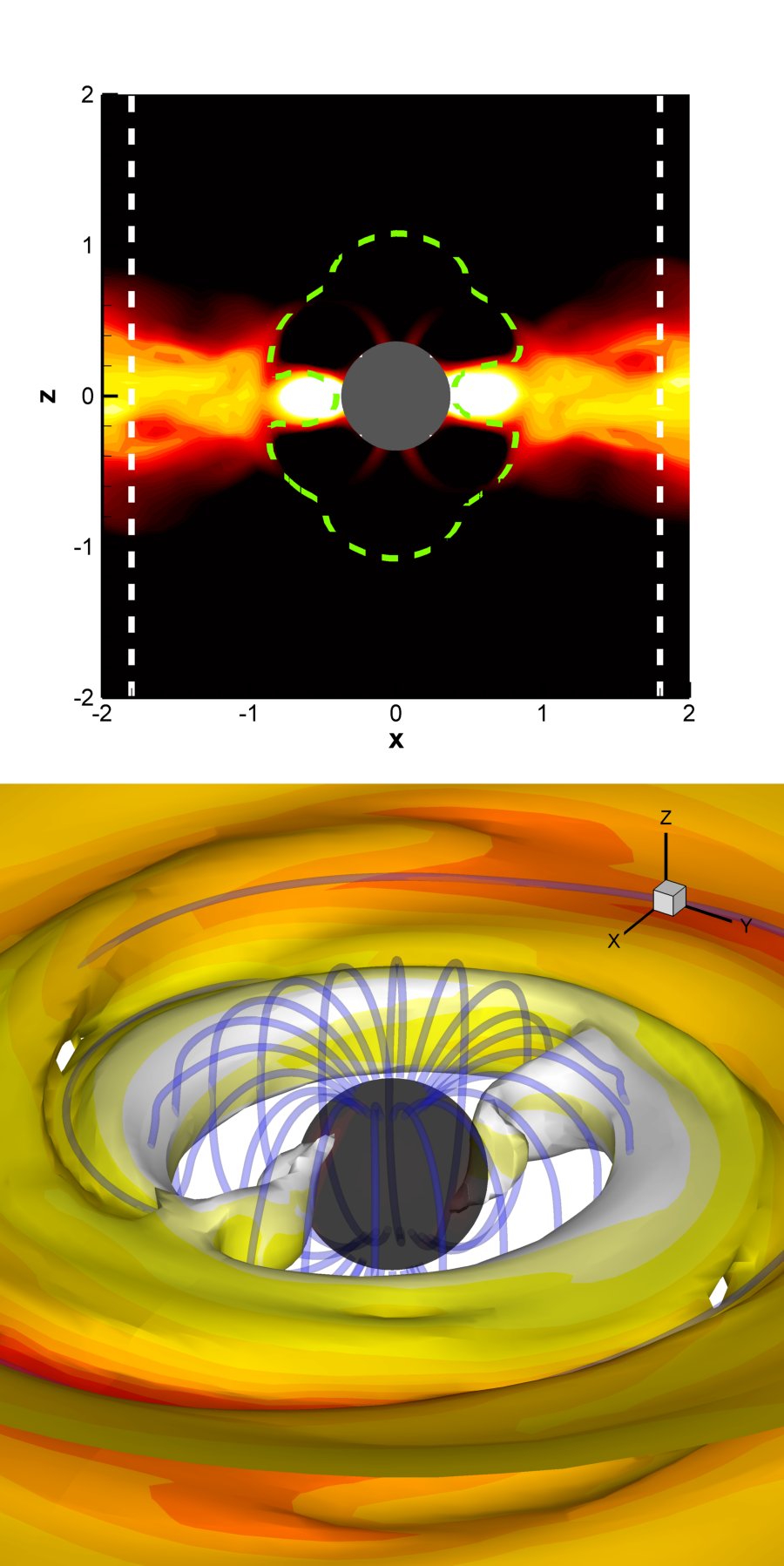}
 \includegraphics[width=4.5in]{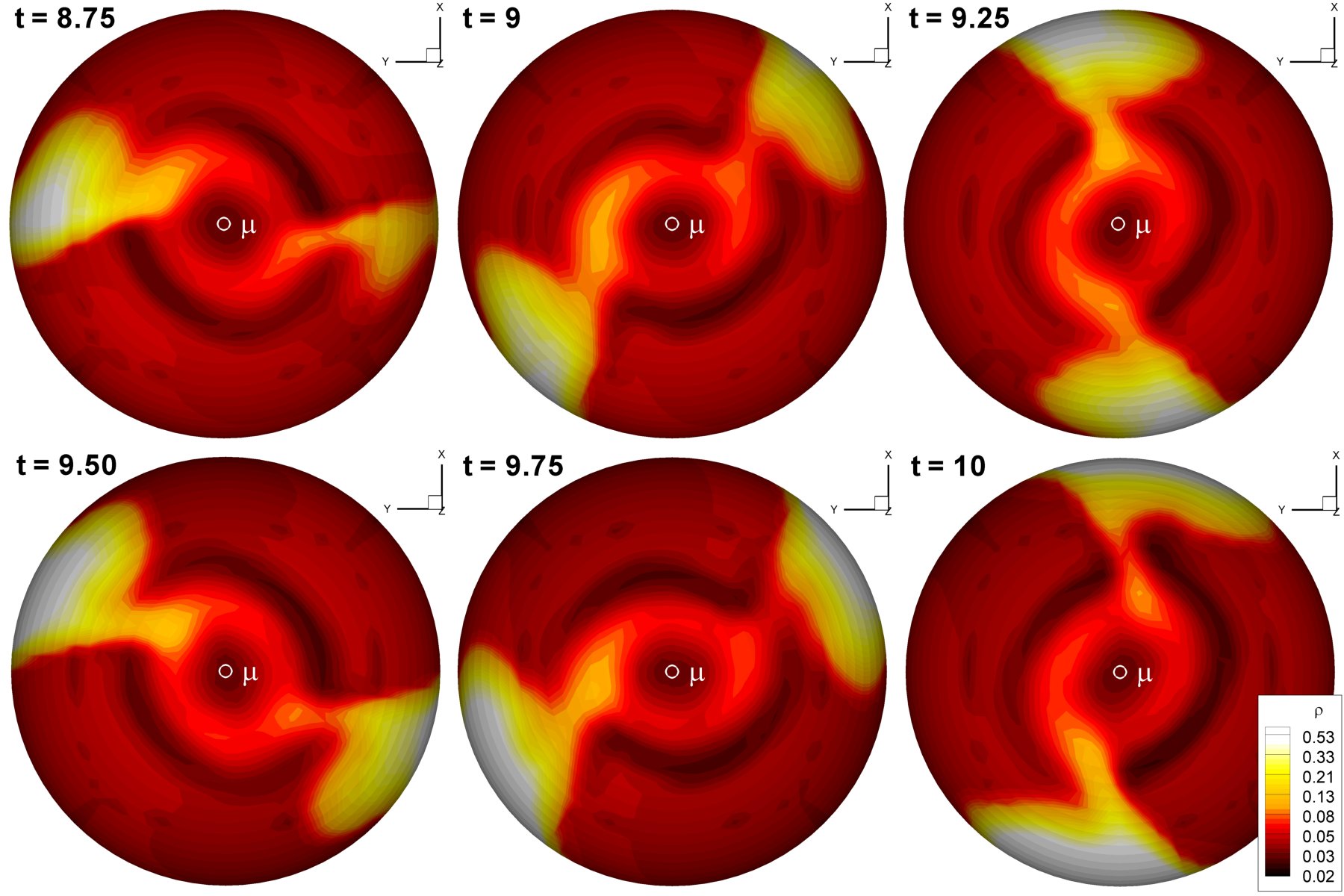}
\end{array}$
 \caption{Position of the hot spots, for the case $\ac=1.8$ ($\spin=5.4$ms),  $\aalpha=0.03$, $\amu=0.3$. This is an example of accretion in the magnetic boundary layer regime. The time difference between the files is $0.25 P_0$. }
 \label{fig:sf-boulay-join}
\end{figure*}

The magnetic field of a NS is usually strong enough to truncate the accretion disk at a certain distance, called inner radius or magnetospheric radius, where the magnetic and material stresses are equal \citep{Ghosh:1979p2434}. In most cases, if the inner radius is comparable with the corotation radius, accretion takes place in a stable fashion: matter is stopped in the disk by the magnetic field and moves towards the magnetic poles with the magnetic field lines acting like an invisible funnel (see \fref{fig:sf-stab-join}, left). Depending on the misalignment angle, the hot spot can either be almost fixed or move in a ring-like zone around the magnetic pole \citep[e.g.][]{Romanova:2003}, with an angular velocity dependent on the Keplerian velocity of the zone of the disk which the matter comes from. For the cases in this paper, the misalignment angle is very small ($\mis=2^{\circ}$) and we observe that the hot spot almost always rotates (e.g., \fref{fig:sf-stab-join}, right three columns). In \fref{fig:spotvsmis} several test cases at various $\mis$ are plotted, showing that the spot movement is less and less pronounced as $\mis$ increases, as expected.

This hotspot movement can be visualized in this way: for a moment let us imagine the star as a perfectly aligned rotator. Matter falling onto the polar caps forms a sort of whirlpool. If density enhancements of the flow are formed (see \fref{fig:sf-stab-join} and also \fref{fig:sf-unstab-join}, bottom left: the density over the funnel is not constant) these density enhancements rotate around the magnetic pole with the angular velocity of the zone of the disk they come from. 

But when the disk comes closer, either for a weak magnetic field or for a high accretion rate, Rayleigh-Taylor instabilities appear at the interface between the disk and the magnetosphere. These instabilities produce long and narrow tongues of matter, that penetrate the magnetosphere and fall towards the equator. Matter spilling through the magnetospheric boundaries can, at this point, fall onto the equator or be diverted by the magnetic field lines to a random point between the equator and the magnetic poles (see \fref{fig:sf-unstab-join}, left; see also \citealt{Kulkarni:2008,Kulkarni:2009}). In this case multiple hot spots appear, whose duration is normally shorter than in the stable case, and whose motion is more complicated, because the tongues of matter originating them follow random paths before reaching the star. The formation of instabilities does not usually prevent the funnel from forming, as \fref{fig:sf-unstab-join} shows. The angular velocities of the unstable spots are measurable from the spot-omega diagram described below and thus, as we are going to see in the next sections, they can give observable effects on the lightcurves.

If the disk comes very close to the surface of the star, matter can penetrate the field lines forming two ordered antipodal tongues which fall towards the equator of the star \citep{Romanova:2009p6603}, as in \fref{fig:sf-boulay-join}. The hot spots produced in this way last longer than in the above cases, and move with a stable angular velocity corresponding to the Keplerian velocity at the truncation radius.

\subsection{The spot-omega diagram}\label{sec:so}
\begin{figure}
 \centering
 \includegraphics[width=3in]{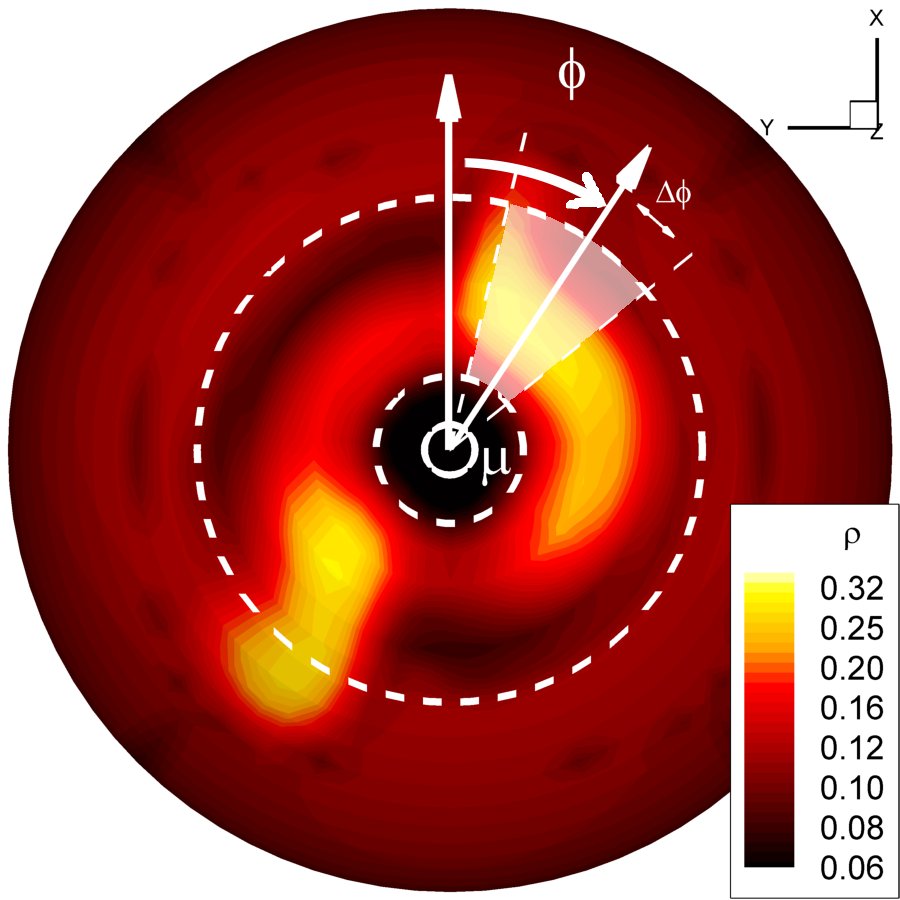}
 \caption{The contour levels show the density distribution on the star's surface, as seen from the magnetic pole. To draw the spot omega diagram, we choose a ring around the magnetic pole, delimited by the dashed lines. Then, for every direction $\phi$ we plot the average density in the area delimited by the ring borders and the directions $\phi+d\phi$ and $\phi - d\phi$, obtaining a plot like the bottom of \fref{fig:so-wlet-join} }
 \label{fig:spotpos}
\end{figure}
We need a tool to describe the movement of the hot spots on the surface of the star in a quantitative way. A tool which not only highlights the movements, but also gives us a way to measure their angular velocity and their duration.

The spot-omega diagram, first introduced in \citet{Kulkarni:2009}, shows the variation of the azimuthal density distribution around the magnetic pole during accretion. 
To obtain it, we choose a ring around the magnetic pole as in \fref{fig:spotpos}, whose limits are two angles $\theta_1$ and $\theta_2$ with respect to the magnetic pole. Then, for every direction $\phi$ around the magnetic pole, we calculate the average density of accreting matter in the area enclosed by $\phi\pm\Delta\phi$ (with $\Delta\phi$ a small step) and the borders of the ring. We do this for every time step of our simulation. This gives us the azimuthal density distribution as a function of time. 

In the resulting diagram, rotations of the spots appear as drifts of the maximum of the flux with time. If we draw a line along the drifts, evidently $\omega=\Delta\phi/\Delta t$, and so the slopes of the curves give us a measure of the angular velocity of the moving spots.

To highlight movements which are longer than one rotation around the star, we can replicate the plot in the vertical direction, going from $\phi=0$ to, say, $\phi=8\pi$ instead of $2\pi$, obtaining for example the plot shown in \fref{fig:so-wlet-join}.

The spot-omega diagram can be analogously used to track the hot spots formed in the equatorial plane, choosing as $\theta_1$ and $\theta_2$ two angles that enclose the equator, measured with respect to the rotational axis instead of the magnetic axis.

\subsection{Measurement of the hot spots' frequency with the spot-omega diagram}\label{sec:proc}
\begin{figure*}
 \centering
 \includegraphics[width=6in]{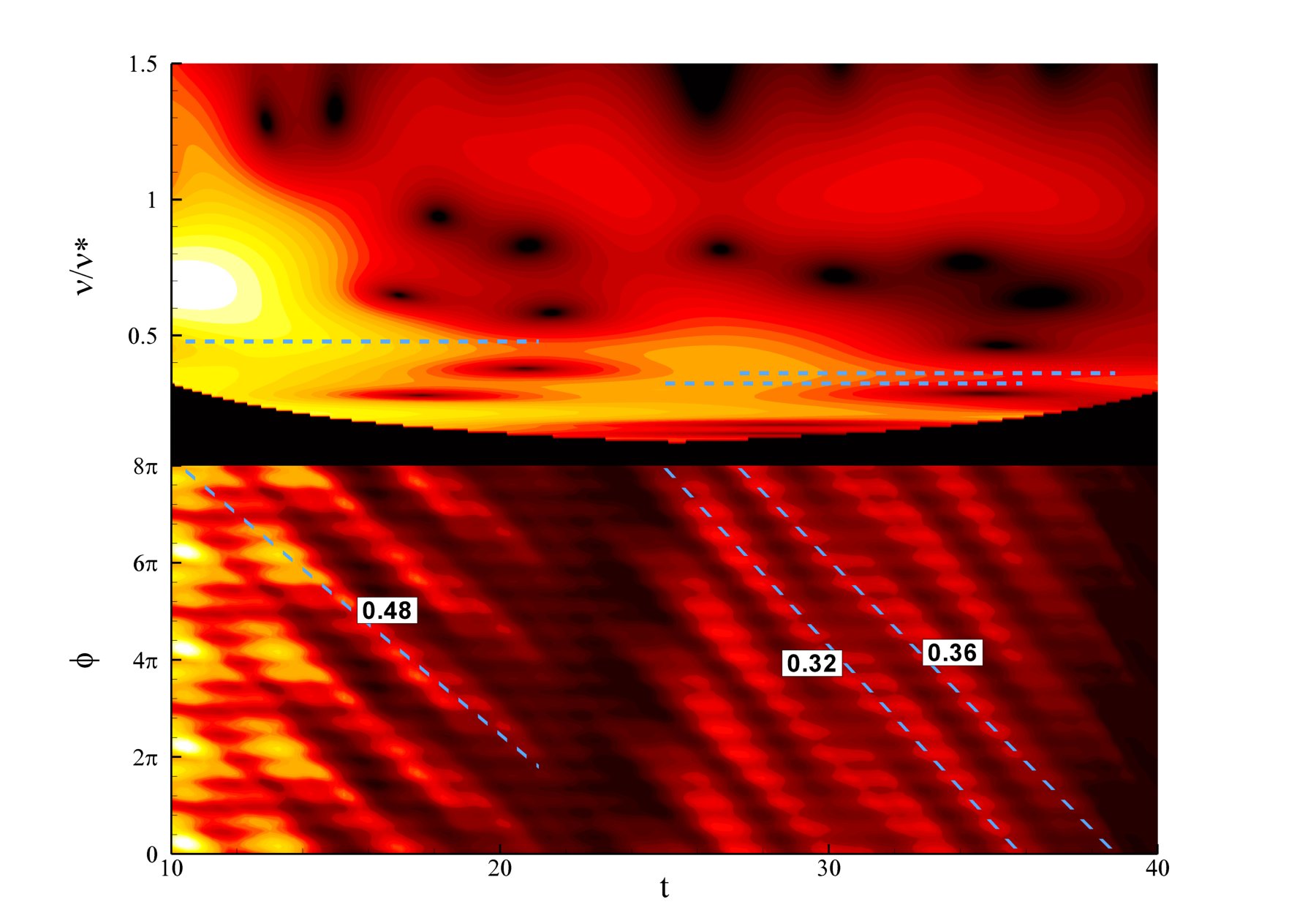}
 \caption{Frequency analysis for $\ac=1.5$,  $\aalpha=0.04$, $\amu=2$, corresponding to a NS with $\spin=4.1$ms and $B=10^9$G. Top panel: wavelet transform of the emission from the surface of the star, as seen by an observer at 45 degrees. Bottom: the spot-omega diagram. The dashed lines lines follow the movement around the magnetic pole of the zones at brightest intensity (``hot spots'') as explained in \sref{sec:so}. Values in the boxes show the angular velocity in units of angular velocity of the star, and the same values are plotted with the blue dashed lines on the wavelet diagram. The horizontal small features in the spot-omega diagram are due to artifacts intrinsic to the cubed-sphere grid for very small misalignment angles.}
  \label{fig:so-wlet-join}
\end{figure*}

 \begin{figure*}
  \centering
 \includegraphics[width=6in]{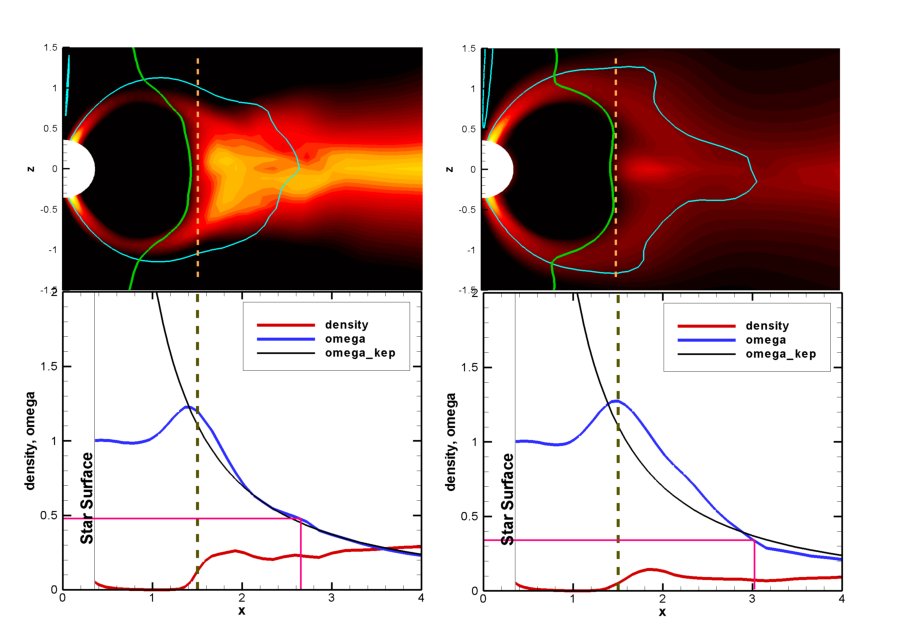}
 \caption{Density and angular velocity distribution around the star at $t=15$ ({\em Left}) and $t=30$ ({\em Right}), in the $xz$ plane ({\em Top}) and along the $x$ direction in the equatorial plane ({\em Bottom}). In the plots at the top, blue contour lines represent the omega distribution, while floods represent the density distribution. The boxes mark the contour lines corresponding to the velocities seen in \fref{fig:so-wlet-join}. The green continuous line represent the magnetic boundary layer (for which matter pressure is equal to the magnetic field pressure) and the dashed line marks the corotation radius. In the bottom plot, the black line marks the radius at which matter has the same angular velocities as in the top plots.}
  \label{fig:xz-e-join}
\end{figure*}
In the spot-omega diagram the movement of the spots is shown by the variation of position of the flux maxima with time.
The angular velocity $\omega$ of the spots is straightforward to calculate from the slopes of the curves the moving spots produce. The values in the boxes are the angular velocities in the frame of reference of a distant observer, in units of $\omega_*$.

Before utilizing the frequencies obtained from the spot-omega diagram, let us justify its use. Our goal is to show that the movements whose frequencies can be observed in the spot-omega diagram also produce oscillations in the lightcurves, and thus traces in the power spectrum.

After obtaining the diagram, we independently calculate the lightcurve produced by the surface of the star. This lightcurve is produced in the approximation of perfect black body emission of all the kinetic energy of the accreting matter and includes corrections for relativistic D\"oppler effect, light bending and time delay \citep{Kulkarni:2005p4093}. 

We analyze it with a wavelet transform, in order to verify if frequencies associated with the spot movement can be extracted by means of spectral analysis. 

Here we show an example of stable accretion onto a neutron star. The parameters chosen for this case are $\ac=1.5$ (corresponding to a NS with $\spin=4.1$ms), $\mis=2\degree$, $\aalpha=0.04$, $\amu=2$.
The results for this case can be summed up as in \fref{fig:so-wlet-join}.

Not surprisingly, we can see a correlation between the slopes observed in the spot-omega diagram and some features of the wavelet transform. Moreover, we should take into account the fact that the typical time scale of our simulations is a few rotations of the star (milliseconds in the case of AMXPs). The wavelet transform only has a very short time to record spectral features. In real accreting systems, time scales for average density changes in the disk are much longer. Therefore, we can reasonably expect to see the emission from the moving spots as spectral features, in contrast with the usual assumptions found in the literature.

As shown in \fref{fig:xz-e-join}, the velocity of the spot is consistent with the Keplerian velocity of the zone of the disk where the funnel starts from.

Thus, so far we have shown that spot motions produce visible features in the power spectra, and can be tracked using the spot-omega diagram. 

\section{Monte Carlo simulation of long-term lightcurves and production of QPOs}\label{sec:mc}
In the last section we showed that during accretion, in both the stable and the unstable cases, spots can move on the surface of the star with a different velocity than the star itself. What is evident from our simulations is that the moving spots are as bright as the ones produced in other simulations, in which the spots were fixed. If the fixed spots are able to produce the strong and stable signal we observe in Accreting Millisecond Pulsars, we believe that the moving ones must be visible through particular features in the power spectrum.

We suggest that QPOs may be among these features. To show this, we start by observing that the spot movements shown in 3D simulations last a very short time, only a few rotations of the star, because disk conditions are changing very fast. In real accretion disks, disk properties change on much longer timescales, say minutes at least. We can safely assume that if movements of spots exist, they will probably last much longer than in simulations. Nonetheless, we can expect the flux of matter not to be constant, and so the spots may be sometimes brighter, sometimes less. 

These bright moving spots emit a signal which, from an observer point of view, must be modulated with a frequency corresponding to the angular velocity of the spots in the observer's frame. Our way to model this kind of emission is to imagine consecutive trains of sinusoids, of variable duration and amplitude but constant or quasi-constant frequency. To simulate the non-pulsed emission from the disk, we add a fixed background. 

Furthermore, we see from simulations that near the $\spinf - \mu$ plane, where the infall of matter is slightly more favorable, the spots are slightly brighter on average. We model this as an additional pulsed signal to add to the rest of the lightcurve. This pulsed signal, coming from a fixed point on the star, is rotating with the star and will thus have the frequency of the star $\spinf$.

At the end of this process, we have a single, continuous lightcurve in which we can separate three components: a fixed background representing the overall average emission from the star, a continuous pulsed signal with the frequency of the star, and a certain number of wavetrains with the frequency of the spots.

As a final icing on the cake, we notice that normally accreting neutron stars are observed in X-rays, and therefore the observed signal is not a continuous lightcurve, but rather a series of photons which are detected as ``events'' by the instruments in X-ray telescopes. To switch from a continuous lightcurve to a more realistic series of events, we take advantage of the fact that common detection rates for this kind of objects in real observations are of the order of $10^{2 - 3} \gamma/s$. For millisecond pulsars, we have approximately one event per rotation, and we can thus safely assume these events as rare. In fact, these events are very well described by a Poissonian variate with an amplitude of probability corresponding to the number of photons detected per unit time.

Then, we obtain a Poissonian random variate with the given amplitude and thus create an event series with the same statistical properties as the ``real'' ones. This helps us compare the results in our Monte Carlo simulation with the observational properties of QPOs from real observations.

\subsection{Fixed wavetrains and QPOs}\label{sec:q}

The movements observed in 3D simulations produce a modulation in the lightcurve emitted by the star. Let us assume that this modulation is purely sinusoidal, but of fixed duration, and thus the product of a sinusoid and a rectangular window function. It can be shown that the Fourier transform of such a sinusoidal wavetrain whose duration is $T$ is a broad peak for which $\Delta \nu \approx 1/ T$, where $\Delta \nu$ is the width of the peak. Hence, if we say that $T$ is $x$ times the period $\tau$ of the impulses, then the quality factor is:
\begin {equation}\label{eq:q}
 Q =\frac {\nu}{ \Delta \nu } = \frac{T} {\tau} = \frac{x\tau}{\tau}  = x.
\end {equation}
Thus, \eref{eq:q} effectively says that the $Q$ factor of a broad peak in the power spectrum produced by short wavetrains gives us an estimate on the duration of the wavetrains themselves. In nature, the $Q$ factor of kHz QPOs is such that, if the model we propose is correct, the spot movements should last for hundreds of rotations in some cases. In our 3D simulations the hot spots rotate with a stable frequency for a much shorter time, but we must consider that the whole simulations only lasts tens of rotations, due to numerical problems that do not permit to maintain stable disk conditions. In nature the disk conditions are, in all probability, much stabler.

\subsection{Description of the model and implementation}
In this section we show in more detail the parameters of our model and the procedure to produce the simulated lightcurve. 
\subsubsection{Random number generation}\label{sec:random}
To generate the random variates we used the very well tested functions of the GNU Scientific Library v.1.10 \citep{iGAL02a}.
As random number generator, we chose the {\em luxury random number} algorithm of L\"uscher ({\tt ranlxd2}), characterized by a period of about $10^{171}$, as implemented in the library itself. 

\subsubsection{Parameters}\label{sec:parameters}
Every moving spot is represented by a wavetrain described by the following parameters:
\begin{itemize}
 \item {\bf spot frequency} $\spotf$, corresponding to the angular velocity of the spot in the observer frame of reference; its value can be constant or vary in a given range;
\item {\bf duration} $\tdur$, whose values can assume a range of values between $\tdurmin$ and $\tdurmax$;
 \item {\bf background} $B$, the amplitude (in $\gamma/s$), constant, of the simulated lightcurve;
 \item {\bf oscillation amplitude} $A$, constant for a given wavetrain, but varying between different wavetrains from $A_{min}$ to $A_{max}$. Its value is expressed as a pulsed fraction of the background. It can also be further modulated as exponentially decreasing or Gaussian shaped during a given wavetrain;
 \item {\bf next spot appearance time} $\tnext$, corresponding to the time between the appearance of two consecutive spots, again varying between $\tnextmin$ to $\tnextmax$;
 \item {\bf phase} $\phi$, which can assume any value between $0$ and $2\pi$.
\end{itemize}

Furthermore, we can also choose the {\bf frequency of the star} $\omega_*$, used to produce the fixed spot emission and also as a unit of time to be consistent with the units used in 3D simulations, and the {\bf background level} $B$, constant during the whole simulation.

\subsubsection{Procedure}\label{sec:mcproc}

\begin{figure}
 \centering
 \includegraphics[width=3in]{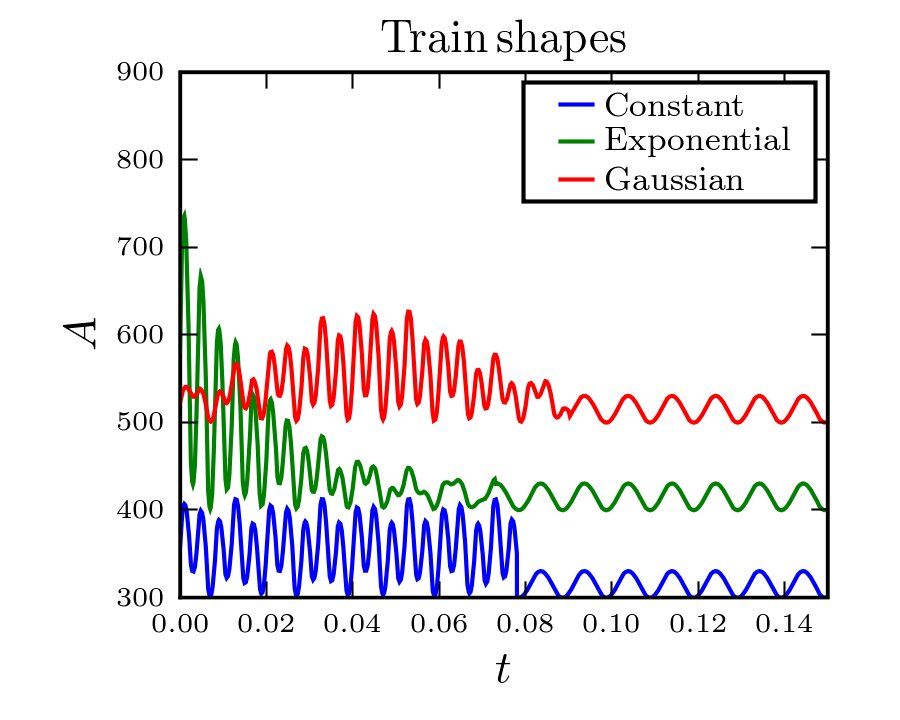}
 \caption{Examples of possible synthetic wavetrains: we can notice the constant sinusoidal modulation with the frequency of the star $\spinf$ and several different behaviors of the intensity of the hot spot, with its modulation at the frequency of the spots $\spotf$.}
 \label{fig:trainshapes}
\end{figure}

We first produce the continuous lightcurve by taking the following steps:
\begin{enumerate}
\item we choose a frequency $\nu$, a sampling time $\tau$ and a total length $L$ of the lightcurve;
\item we choose a pulse shape $\Ps$ (by default sinusoidal) and a window function $\Ws$ (typically flat, Gaussian or exponentially decaying as in \fref{fig:trainshapes}) giving the overall intensity variation of the wavetrains. It is of particular importance to also choose if we want a constant average intensity of the source (and so $\int{\Ps}(t) dt=0$) or if we want the spots to create an additional luminosity (``burst-like'');
\item\label{item:generate} we generate a random value (distributed according to a uniform or Gaussian variate) for $A$ and $\tdur$;
\item we generate a random value (distributed according to a uniform, Gaussian or Exponential decay-like variate) for $\tnext$;
\item we produce a random value for $\phi$;
\item we generate a wavetrain of duration $\tdur$ from $\tnext$, according to the formula:
\begin{equation}
  a(t)= A \Ps (t+\phi\spott) \Ws(t)
\end{equation}
\item we go back to~\ref{item:generate} if $t < L$
\item to the lightcurve obtained, we add a constant background and a continuous sinusoidal modulation with frequency $\omega_*$.
\end{enumerate}
At this point, we have the complete simulated lightcurve as it would be seen by an optical telescope. We only need the following few steps to obtain a series of events as in real X-ray observations:
\begin{enumerate}
\item for every $t$ we produce a random integer number $N_{events}$, distributed according to a Poissonian variate with amplitude $\tau A(t)$, corresponding to the expected number of events in the time $\tau$;
\item we write in the output file the number of events $N_{events}$ ``recorded'' at the time $t$.
\end{enumerate}

\subsection {Spectral analysis}\label{sec:fft}
Fourier Transform is the main technique used to find periodicities in time series. The Fourier transform, in fact, gives a representation of the input function in term of its harmonic content. With Fourier transforms it is possible to find periodicities in a time series even if the amplitude of the periodic component is very small. 

We performed a Fast Fourier Transform\footnote{There are several algorithms to calculate Fourier transforms of sampled data. The fastest ones, which exploit the properties of Fourier transforms and also of the data representation in computers, are known as Fast Fourier Transforms.} on the data by means of the FFTW3 library \citep{FFTW05}, and obtained the power spectrum according to the formula
\begin{equation}
P_j=\frac{x^2 + y^2}{N}
\end{equation}
where $x$ and $y$ are the real and imaginary parts, respectively, of the Fourier transform of the time series, and $N$ is the total number of simulated events. This is the standard normalization proposed by \citet{1983ApJ...266..160L}, which is particularly useful in the case of Poissonian events. In fact, using this normalization the power spectrum of a constant background follows a $\chi^2$ distribution with $2N_{bin}$ degrees of freedom, where $N_{bin}$ is the total number of bins in the spectrum. In this way we have a very solid statistical framework to help us interpret our results.

In order to improve the signal to noise ratio, we performed the FFT on M consecutive chunks of data and obtained a single FFT from the average of the chunks. With the chosen normalization, the background has an average of 2 and a standard deviation of $2/\sqrt{M}$ \citep{1988tns..conf...27V}. 

\subsection{Example of Monte Carlo-produced QPO}\label{sec:mcres}
\begin{figure*}
 \centering
$\begin{array}{cc}
 \includegraphics[width=3in]{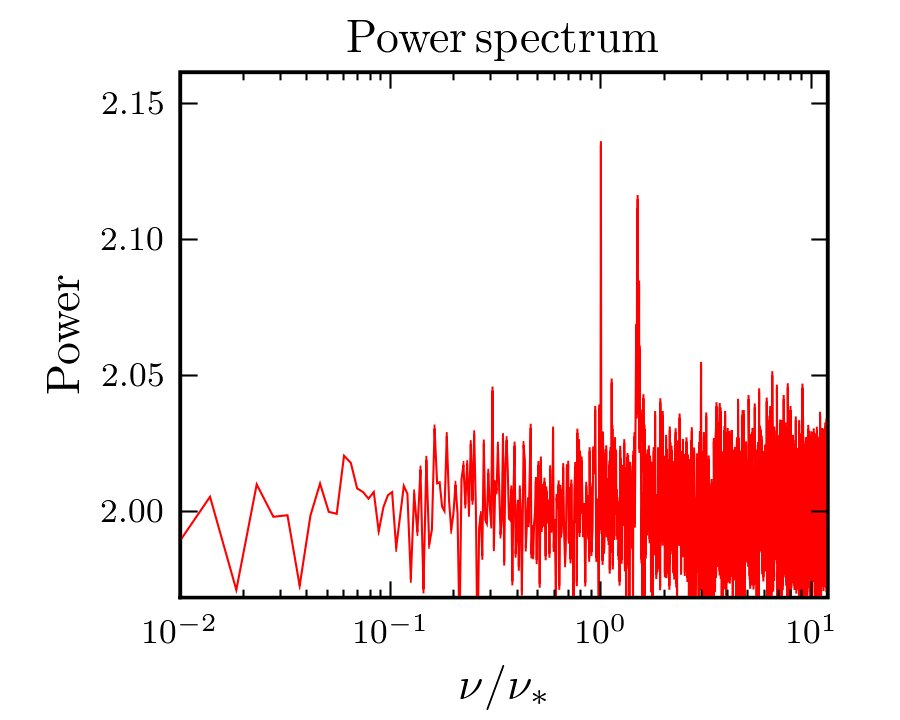}
 \includegraphics[width=3in]{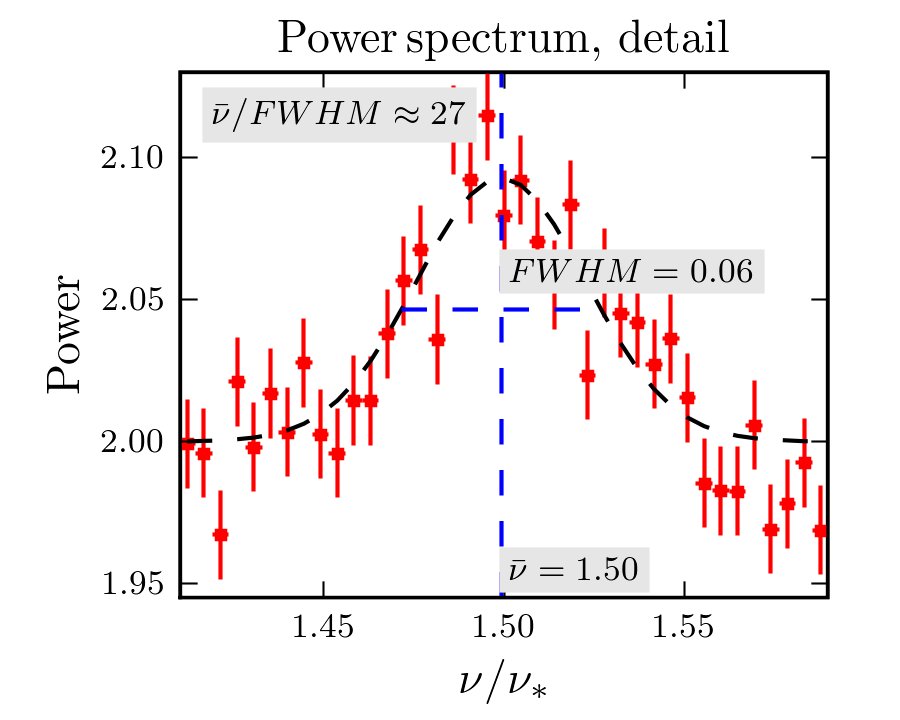}
\end{array}$
 \caption{({\em Left}) Spectrum of the simulated time series, with $\spin=3$ms,  $\spott=2$ms, $\tdur=50$ms, $\tnext=100$ms, $A=20\%$, $B=300\gamma/s$. In this plot we can observe the QPO at $\nu=1.5\spinf$ and the peak at the frequency of the star ($\nu=1$). The length of the simulated time series was $10000$s and the spectrum was obtained averaging the spectra of 16384 subintervals. ({\em Right}) Gaussian fit of the simulated QPO}
 \label{fig:spectrum}
\end{figure*}

The case in \fref{fig:spectrum} is corresponding to a spin period $\spin=3$ms and a hot spot period $\spott=2$ms; we used a duration of the spot motion of $\tdur=50$ms and $\tnext=100$ms. The pulsed fraction of the moving spots was $20\%$, while that of the fixed spot was $10\%$. We can see the QPO peak at $1.5\spinf$ and the sharp peak at the frequency of the star.

On the right hand side of the same figure, the QPO peak is fitted and some details are provided: as one can see, the peak is well fitted by a Gaussian with $\sigma=0.01\spinf$.

\section{Correlations with the mass accretion rate}\label{sec:mdot}

\begin{figure*}
 \centering
$\begin{array}{cc}
 \includegraphics[width=3in]{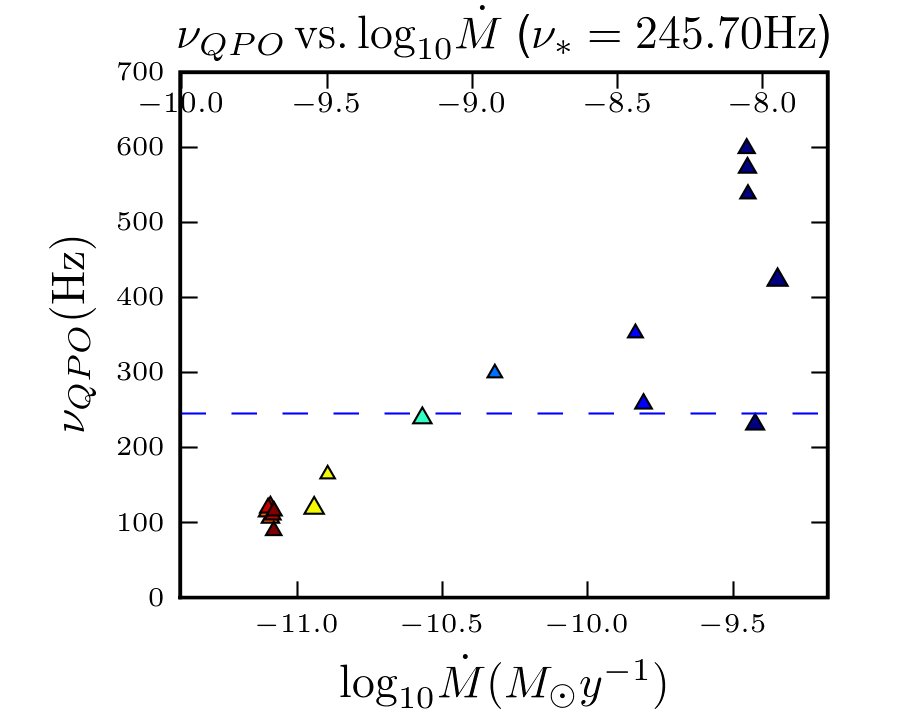} &
 \includegraphics[width=3in]{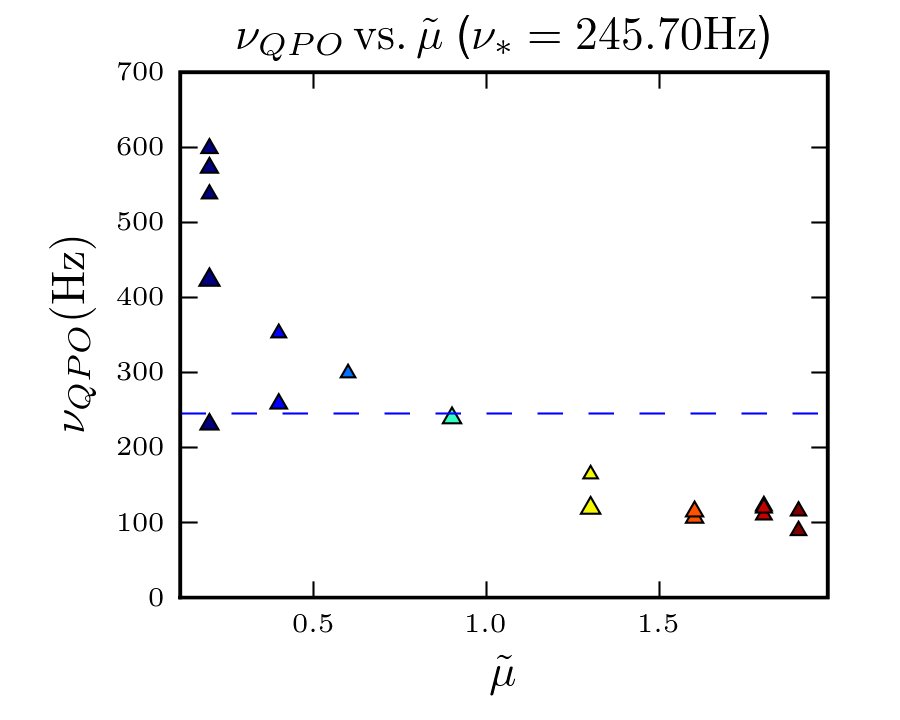}\\
\end{array}$
 \caption {Left: correlation between the frequencies $\nuq$ observed in the spot-omega diagram and the mass accretion rate for the case $\ac=1.5$ (\spin=4.1ms), $\aalpha=0.03$. The scale at the top is for a neutron star with $B=5\cdot10^8$G, the one at the bottom for $B=10^8$G. Right: correlation with the dimensionless parameter $\amu$ in the same case. The dashed line marks the frequency of the star. The colors indicate $\amu$, in order to compare the position of the points in the two plots, and the size of the mark is proportional to the time of appearance of the features in the simulation. }
 \label{fig:mdot15}
\end{figure*}

\begin{figure*}
 \centering
$\begin{array}{cc}
 \includegraphics[width=3in]{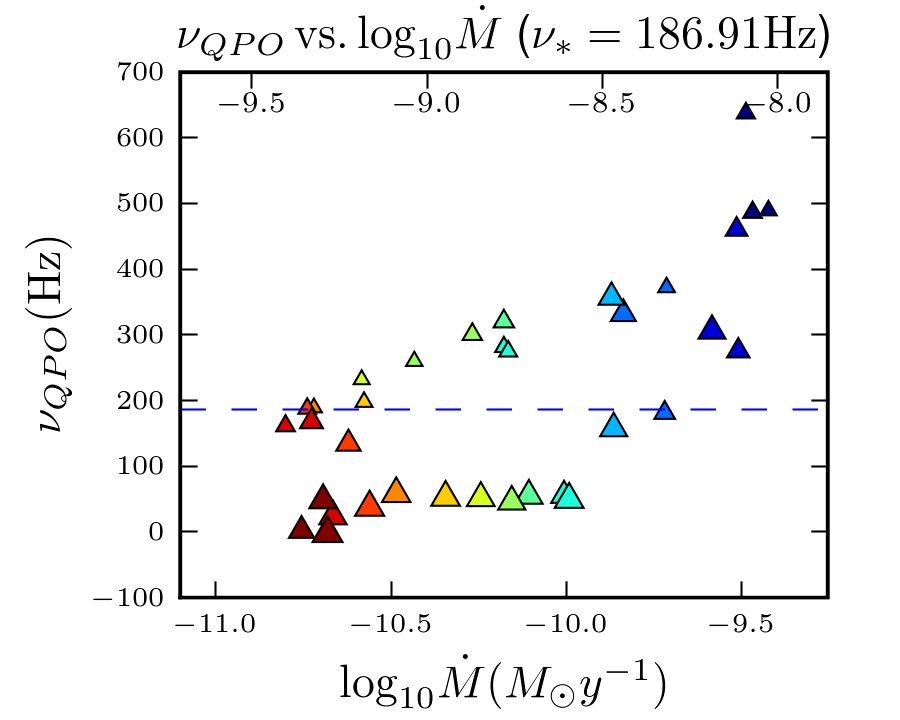} &
 \includegraphics[width=3in]{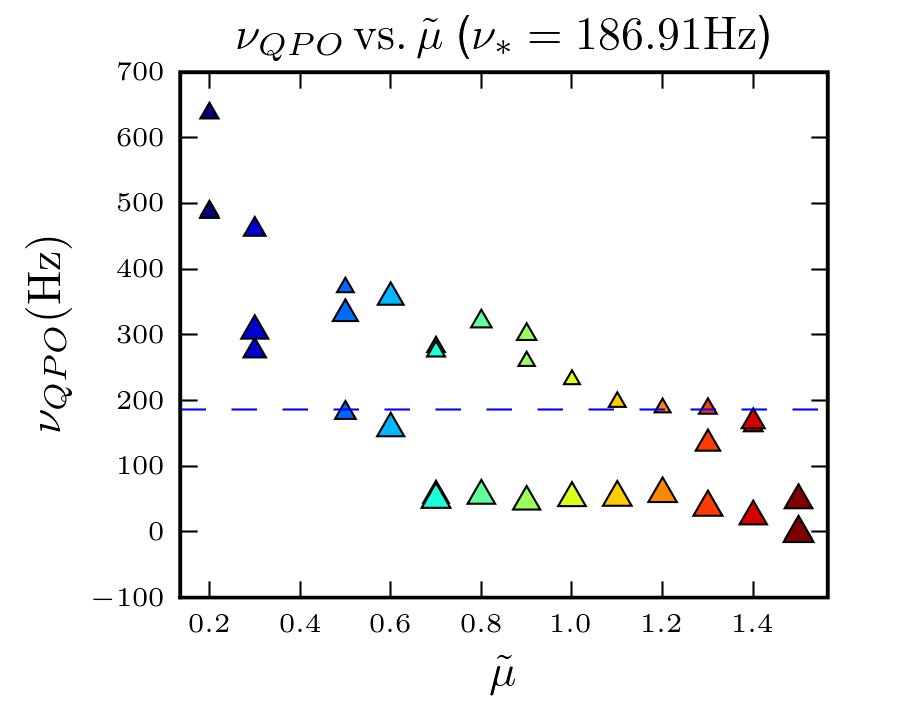}\\
 \includegraphics[width=3in]{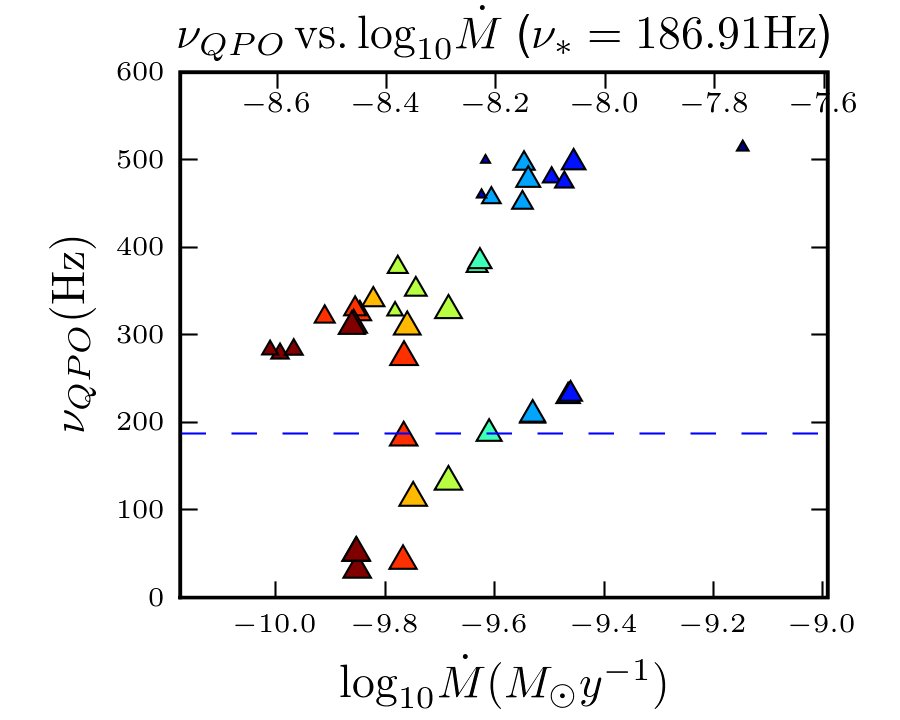} &
 \includegraphics[width=3in]{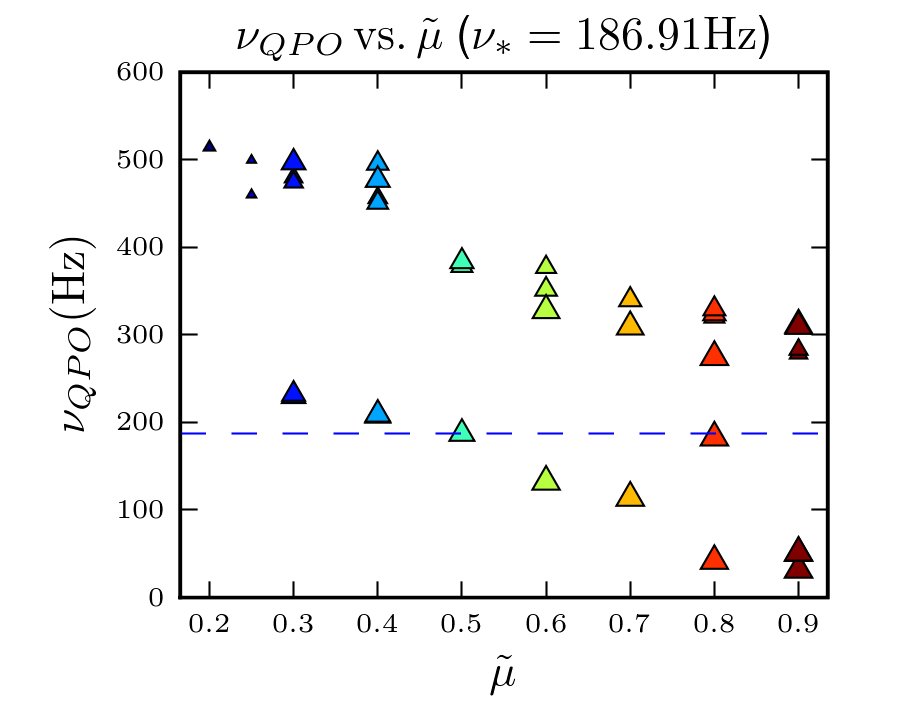} \\
 \includegraphics[width=3in]{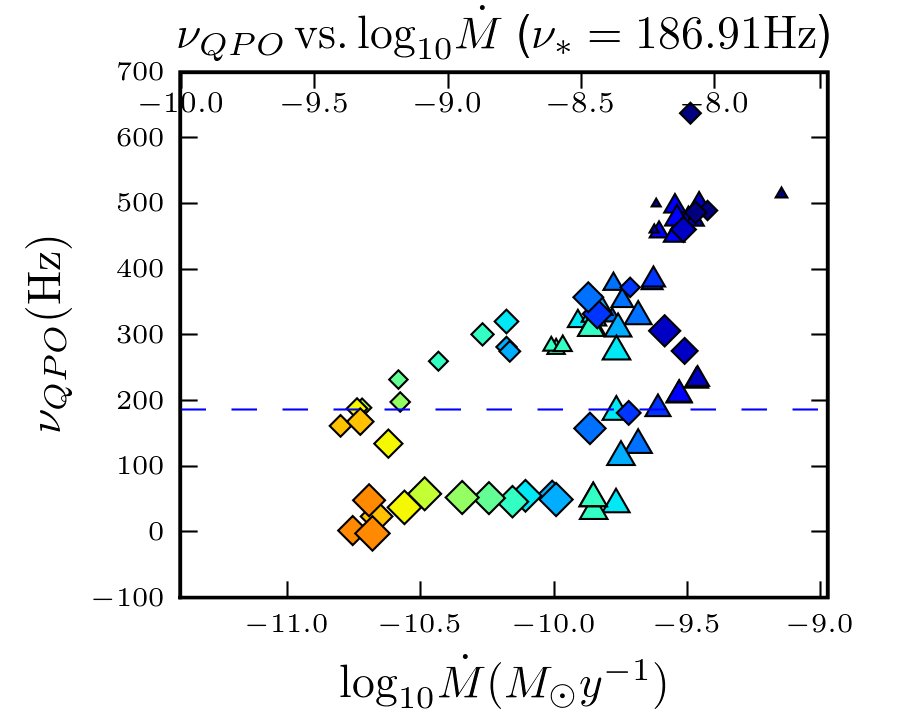} &
 \includegraphics[width=3in]{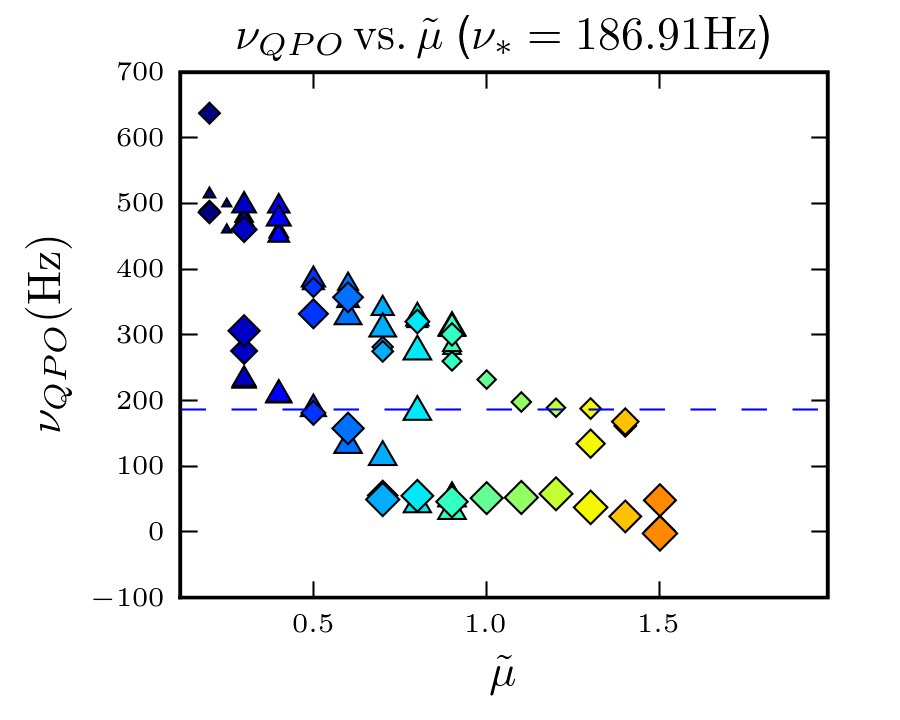}
\end{array}$
 \caption {$\nuq$ variation with respect to $\mathrm{log}_{10}\dot{M}$ and $\amu$ for $\ac=1.8$ (\spin=5.4ms). As before, the scale at the top of the ``$\nuq$ vs. $\dot{M}$'' plot is for a neutron star with $B=5\cdot10^8$G, the one at the bottom for $B=10^8$G. Top and center cases have $\aalpha=0.02$ and $\aalpha=0.03$ respectively, while at the bottom the two cases are plotted simultaneously.}
 \label{fig:mdot18}
\end{figure*}

In \sref{sec:3d} and~\ref{sec:mc} we showed that movements of the hot spots can produce QPOs. In this section we show that those movements are not a fortuitous event, but in fact they present a predictable behavior. To show this, we study the change on the angular velocity of the hot spots as the mass accretion rate changes, performing a number of new simulations with different values of $\amu$ using \eref{eq:mdot} to find the corresponding values of $\mdot$. We then represent the frequencies observed from the spot-omega diagrams against the corresponding values of $\amu$ and $\mdot$ in plots like those in Figs~\ref{fig:mdot15} and \ref{fig:mdot18}. To be complete, we include in the study all the frequencies visible in the spot-omega diagrams for all cases. We do not perform any choices on the frequencies to plot. Multiple frequencies appear in some cases simultaneously, in others at different times. The size of the bullets in the plots is proportional to the time the plotted features appear at during the runs. In this way is possible to see whether multiple features appear together or not.

\begin{figure}
 \centering
 \includegraphics[width=3in]{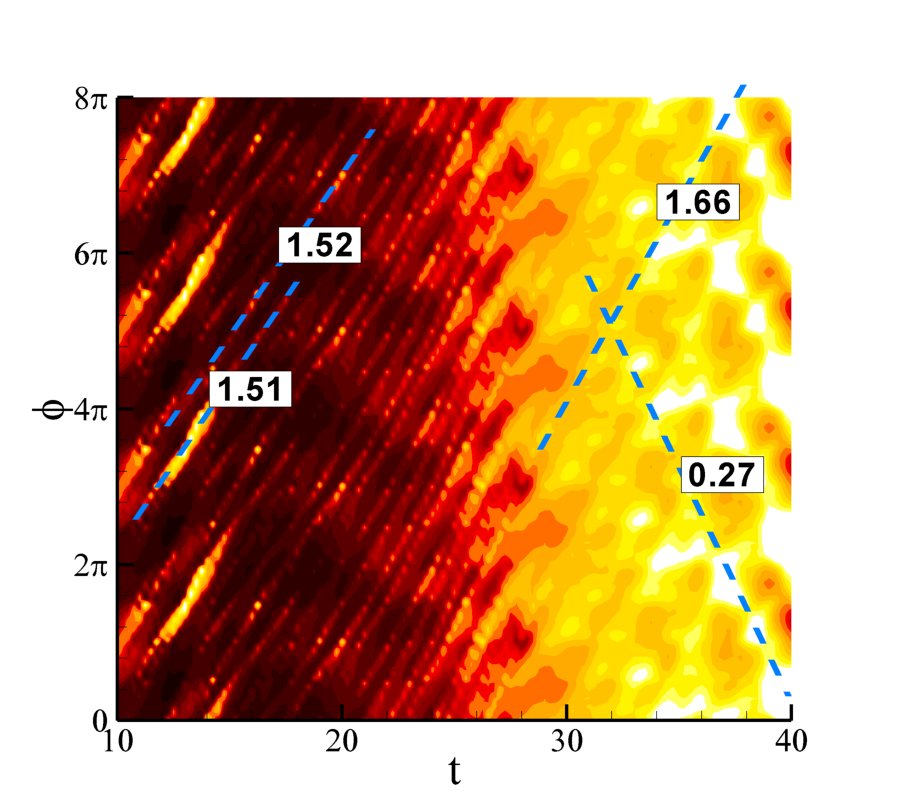}
 \caption{Spot-omega diagram for the case  $\ac=1.8$ (\spin=5.4ms), $\aalpha=0.03$, $\amu=0.9$. In this case is possible to notice the coexistence of two motions of the hot spots, with two different scales of velocity.}
 \label{fig:doubleso}
\end{figure}

In Figs. \ref{fig:mdot15} and~\ref{fig:mdot18} the $\nuq$s obtained from the spot-omega diagrams are plotted against the mass accretion rate and the dimensionless parameter $\amu$ (which determines the size of the magnetosphere) for several cases. As previously said, we choose to model a neutron star with $B=10^9$G, $R=10$km, $M=1.4M_{\odot}$. According to \eref {eq:mdot} we can use $\amu$ to change $\mdot_0$, investigating the variation of $\nuq$ with respect to the measured value of $\mdot$.

The case in \fref{fig:mdot15} describes an NS with $\spin=4.1$ms. This is a case of almost stable accretion, except for high values of the accretion rate. As the accretion rate grows, the magnetosphere becomes smaller and smaller, and the hot spots move around the magnetic pole with an increasing frequency, as expected. When $\mdot$ reaches higher values the accretion is not stable anymore, and multiple hot spots appear, originating the multiple frequencies observed in \fref{fig:mdot15}. In this case all the simulations were short compared to the ones in the next paragraph because of the appearance of numerical instabilities, and so all frequencies are observed at an early time.

The case in \fref{fig:mdot18} describes unstable behavior, like we can see in \fref{fig:sf-unstab-join}. In this case we can observe a peculiar behavior: the hot spots in the ring around the magnetic pole and the ones produced by instabilities move with different angular velocities. This movement is visible also from the spot-omega diagram in \fref{fig:doubleso}. This produces two traces in the spot-omega diagram, whose frequencies present the same approximate behavior, as $\mdot$ changes, as the ones seen in the first case. The frequency separation of the two peaks is in the correct range of values ($200 - 300$Hz) to suggest that the twin kHz QPO peaks often observed in nature could be originated in a similar way. A confirmation of this idea can be the fact that the upper kHz QPO is usually much less coherent than the lower one (\citealt{diSalvo:2003p5128}; \citealt{Barret:2006p7645} investigate this particular aspect for a wide sample of LMXBs). In our simulations, the ``upper'' peak is produced by instabilities while the ``lower'' one is produced by the motion around the magnetic pole. Normally, unstable accretion is characterized by a much smaller coherence than the stable drift around the magnetic pole, except in the strongly unstable regime in which the two antipodal tongues maintain a constant angular velocity for a relatively long time \citep{Romanova:2009p6603}. 

Both frequencies relate to the zone near the inner disk. The instabilities form at the edge of the disk, and hotspots produced by them have higher frequencies than those of the hotspots in the polar region, which are produced by matter captured by the field lines at a slightly larger distance (see \fref{fig:xz-e-join}, \fref{fig:sf-unstab-join}). The two frequencies move together as $\mdot$ changes because both of them depend on the distance of the disk from the star.

\section{Discussion}
We developed a model of quasi-periodic oscillations based on the spot movements on the surface of the accreting neutron star with a slightly misaligned dipole magnetic field.  3D MHD simulations have shown that spots may rotate faster/slower than the star in cases of stable or unstable accretion  \citep[see also][]{Romanova:2003,Romanova:2004p4095,Kulkarni:2008,Kulkarni:2009} and in the magnetic boundary layer regime \citep{Romanova:2009p6603}. We have shown that in all above cases, the spot movements develop ordered oscillations in the light-curves. Modeling these oscillations as short-lived sinusoidal trains, we showed that if they are powerful enough to be seen, their fingerprint is a QPO. Given the range of values of the frequencies involved, we hypothesize that kHz QPOs could be produced in this way. 

There are several features of QPOs that can give confirmation to this model. First of all, the general trend with the mass accretion rate, as shown in \sref{sec:mdot}, is as it is expected. Though it is not generally true that luminosity and mass accretion rate are proportional to each other, we can assume it to be true on a given source and for a short timescale. This is the case of the simulations in this paper, where only one parameter (the mass accretion rate) changed between the various simulations.

The rms amplitude of real life QPOs has a strong dependence on energy \citep {Berger:1996p4473}, suggesting that the mechanism of emission is different from that of the background. In our opinion, this is because the QPOs are produced on the surface, or perhaps just above the surface by means of shocks in the accretion column similarly to what happens in AMXPs \citep[see for example][]{Gierlinski:2002p5829},  while most of the background is produced by the disk. The rms amplitude is also generally higher in atoll sources (up to 20\%) than in Z sources (2-5\%) \citep{Jonker:2001p4470}. The reason for this difference is probably related to the higher accretion rate of Z sources. 

In some cases  3D simulations show that  two frequencies appear at the same time. This numerical discovery is important for possible interpretation of two QPO peaks observed in a number of LMXBs. In fact, this mechanism for the production of multiple QPOs is able to explain more than the mere appearance of two peaks in the Power Spectrum. When double QPOs appear in observations, they normally present different behavior as regards their $Q$ factor \citep{diSalvo:2003p5128,Barret:2006p7645}, and hence (according to our assumption) their coherence time with the upper peak generally less coherent; in our model, this is due to the fact that the lower one comes from a smooth motion around the magnetic pole, while the upper one is produced by instability tongues, shorter-lasting and thus with a lower coherence. The term ``unstable'' should not mislead the reader though: the unstable regime is very common in simulations, expecially for fast rotators, also at small values of the mass accretion rate, as shown in \fref{fig:mdot18}. A higher coherence of the upper kHz QPO, in our model, could in principle be produced by a very high accretion rate and the onset of the magnetic boundary layer regime, in which instabilities produce long-lasting opposite tongues in the equatorial plane. During this regime though, it is very probable that the zone around the surface of the neutron star is optically thick. Just as observations show, multiple QPOs can appear simultaneously or at different times. 

As regards the models predicting the upper limit of $\nu_u$ as the Keplerian frequency at the ISCO \citep{Miller:1998p4107}, it would be interesting to investigate the matter further. For the values of mass, magnetic field and radius used in the present job, the ISCO is never reached by the inner disk during QPO production with the two channels described because of the onset of the magnetic boundary layer regime. Further investigation should include the use of more extreme values of the mass in order to study the behavior of the inner disk when the ISCO is considerably far from the surface.

Another interesting point is the $\nu-L_x$ correlation observed in many papers \citep[for example][]{Yu:1997p4754,Mendez:1998p6564}. The correlation only holds for a given source and on time scales shorter than a few hours. The main reason why it does not hold across sources can easily be found in the magnetic field: for a given value of the mass accretion rate, different values of the magnetic field correspond to a different size of the magnetosphere and, as a consequence, to different values of the frequency of the moving spots. Again, this is evident from our simulations: the exact same frequency in figure \ref{fig:mdot15} is obtained for $\mdot\approx4\cdot 10^{-11}\msun y^{-1}$ in a system with $B=10^8$G and for $\mdot=10^{-9}\msun y^{-1}$ in a system with $B=5\cdot10^8$G. But the simulations presented here show only one of the many possible configurations of the disk around the star. For a given source, it is possible that changes in the structure (density distribution for example) of the disk produce a different behavior of the moving hot spots, even with the same mass accretion rate. Besides, the relationship between luminosity and mass accretion rate is not clear. This point will be investigated more deeply in future papers, by simulating systems similar to the one shown here but with different initial disk conditions.

The results presented here are based on simulations of neutron stars with very small misalignment angles. An equivalent discussion can be made at slightly larger misalignment angles, where the behavior of the spots does not change considerably (see \fref{fig:spotvsmis}, but also \citealt{Kulkarni:2008,Kulkarni:2009,Romanova:2008}), and obtain almost the same features in the power spectrum. In this case though, from simple geometrical considerations it is reasonable to expect some kind of modulation at the frequency of the star. The least we can expect is a broad Lorentzian-like feature at the frequency of the star \citep{Burderi:1993p328,Lazzati:1997p8008,Menna:2002p949} just from the action of the rotating spot in the polar region, due to the rotation of the star. Its quality factor $Q$ would be comparable to that of the moving hotspots ($Q_*/Q_{spot} \approx {\spinf/\nu_{spot}}$, according to the considerations of \sref{sec:q}). The observation of such a feature, even very dim, in observations where the lower kHz QPO (the one from the funnel flow in this model) is present, would give interesting evidence in favor of this model.

As a final remark, we notice that the difference between the frequencies of the two QPOs is $200-300$Hz, close to the frequency of the star. It is generally known that for relatively slow rotators the frequency difference of QPOs is similar to the frequency of the star. Until about two years ago, it seemed that the $\Delta\nu$ was approximately equal to $\spinf$ for slow rotators and $0.5\spinf$ for fast rotators. \citet{Mendez:2007p5756}, analyzing the matter in detail, find that this distinction is not so clear, and that data can be interpreted and fitted equally well by values of $\Delta\nu$ around $300$Hz for all sources, with the caveat \citep{vanStraaten:2005p4594} of a few AMXPs showing all variability components (and thus also $\Delta\nu$) shifted down by a factor of $\sim 1.5$. The range of values of $\Delta\nu$ found in this work are compatible with both hypotheses, because the simulated system is a slow rotator and $\Delta\nu$ is both near the frequency of the star and $300$Hz. We are planning to investigate the matter with simulations of faster rotators. Nevertheless, this mechanism for the production of double QPOs does not need the two frequencies to be origined from one another, giving a fixed value of $\delta\nu$, which was the main caveat of a somewhat similar model, the sonic point beat frequency model by \citet{Miller:1998p4107}.

From what we observe in this work, the two QPOs are two separate effects of the interaction between the rotating magnetosphere and the disk. The funnel flow comes from a zone just behind the inner radius where the magnetic field is strong enough to capture matter and make it drift towards the pole. The instabilities form right at the inner radius, where magnetic field energy density and ram pressure are equal (the green line in \fref{fig:sf-stab-join}). The results of this paper can give insight into the matter: the frequencies found in observations are a probe of the interaction between the magnetic field and the disk.
 
\section*{Acknowledgements}
MB, LB and TD were supported by the contract ASI-INAF I/088/06/0 for High Energy Astrophysics. MB was also supported by a grant from INFN. MMR and AK  were supported in part by NASA grant NNX08AH25G and by NSF grant AST-0807129. The simulations described in this paper have been perfermed using the NASA's Columbia and Pleiades, and CINECA's BCX, supercomputers, the latter under the 2008/2010 INAF/CINECA agreement. MB thanks the Cornell Plasma Astrophysics Group for the exquisite hospitality they offered and for the use of the 3D code.

\appendix
\section {On the behavior of the spots at different misalignment angles}\label{sec:mis}

\begin{figure*}{ht}
 \centering
 \includegraphics[width=5in]{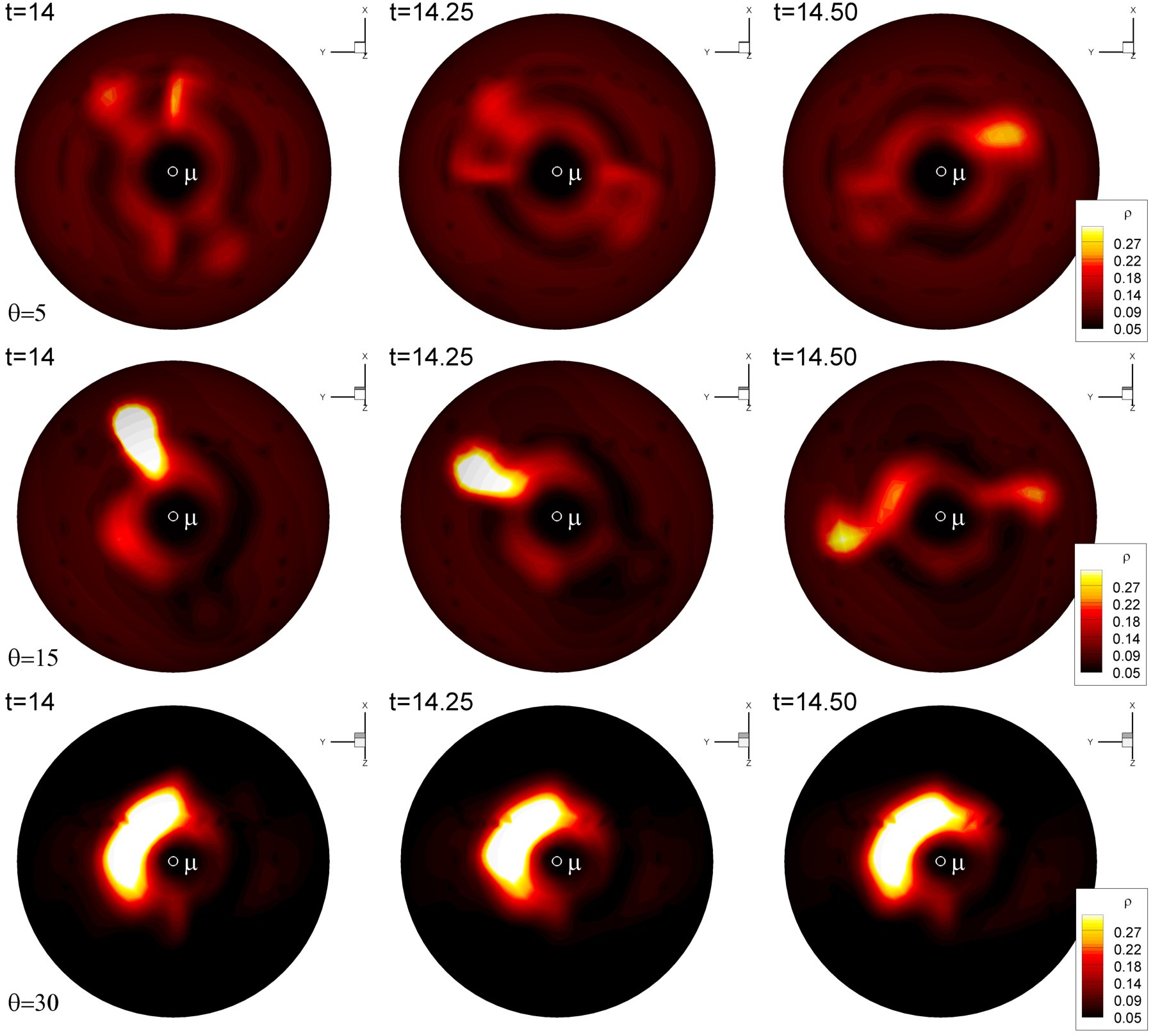}
 \caption{Hot spot shape for $\ac=1.8$ ($\spin=5.4$ms),  $\aalpha=0.03$, $\amu=0.5$, and ({\em Top}) $\theta=5\degree$, ({\em Middle}) $\theta=15\degree$, ({\em Bottom}) $\theta=30\degree$. For small misalignment angles, the hot spot moves with an approximately constant intensity around the magnetic pole. For bigger angles, the hot spot tends to stay fixed or to vary its intensity from a maximum value in the preferred position (which depends on the misalignment angle and on $\mdot$) to a minimum value at the opposite side with respect to the magnetic pole.}
 \label{fig:spotvsmis}
\end{figure*}

In \fref{fig:spotvsmis} we can compare the behavior of the hot spots at different misalignment angles. As can be seen in the picture, the spot movement is not limited to the case with very small misalignment angles, being present even for $\mis=15\degree$. 

\end{document}